\documentclass[letterpaper, 10pt]{article}

\usepackage[letterpaper,bindingoffset=0.2in,%
            left=.8in,right=.8in,top=1in,bottom=1in,%
            footskip=.25in]{geometry}

\usepackage{amsthm}
\usepackage{graphicx}
\usepackage{amsmath}
\usepackage{amssymb}
\usepackage{graphics}
\usepackage{latexsym}
\usepackage[dvips]{epsfig}
\usepackage[nospace]{cite}
\usepackage{subfigure}
\usepackage{tabularx}
\usepackage{booktabs}
\usepackage{color}

\newtheorem{Definition}{Definition}
\newtheorem{Lemma}{Lemma}
\newtheorem{Corollary}{Corollary}
\newtheorem{Proposition}{Proposition}
\newtheorem{Theorem}{Theorem}

\newtheorem{Example}{Example}
\newtheorem{Remark}{Remark}

\def\Pr{{\mathbb{P}}}
\def\E{{\mathbb  E}}

\def\Tr{{\rm {tr} }}

\def\R{{\rm {R} }}

\allowdisplaybreaks[1]
\begin{document}
%
\title{\bf Strong Converse Theorems for Multimessage Networks with Tight Cut-Set Bound}

 \linespread{1.2}
\date{}
\author{\textbf{Silas~L.~Fong}$^{\,a,\,\ast}$ and \textbf{Vincent~Y.~F.~Tan}$^{\,b,\,\ast\ast}$\\ \\
\itshape
$^a\,$Department of Electrical and Computer Engineering, \\
\itshape University of Toronto, Toronto, ON M5S 3G4, Canada \\
 \itshape $^b\,$Department of Electrical and Computer Engineering,\\
\itshape Department of Mathematics,\\
\itshape National University of Singapore, Singapore 117583 \\
\textit{e-mail}: $^\ast\,$\texttt{silas.fong@utoronto.ca}, $^{\ast\ast}\,$\texttt{vtan@nus.edu.sg}
}

\maketitle

\begin{abstract}
This paper considers a multimessage network where each node may send a message to any other node in the network. Under the discrete memoryless model, we prove the strong converse theorem for any network whose cut-set bound is tight, i.e., achievable. Our result implies that for any fixed rate vector that resides outside the capacity region, the average error probabilities of any sequence of length-$n$ codes operated at the rate vector must tend to~$1$ as~$n$ approaches infinity. The proof is based on the method of types and is inspired by the work of Csisz\'{a}r and {K\"{o}rner} in 1982 which fully characterized the reliability function of any discrete memoryless channel (DMC) with feedback for rates above capacity. In addition, we generalize the strong converse theorem to the Gaussian model where each node is subject to an almost-sure power constraint. Important consequences of our results are new strong converses for the Gaussian multiple access channel (MAC) with feedback and the following relay channels under both models: The degraded relay channel (RC), the RC with orthogonal sender components, and the general RC with feedback.
\end{abstract}

\section{Introduction}\label{introduction}
This paper considers a general multimessage network which may consist of multiple nodes. Each node may send a message to any other node in the network. Under the discrete memoryless model where all the input and output alphabets are assumed to be finite, the network is referred to as \emph{discrete memoryless network (DMN)} \cite[Ch.~18]{elgamal}. A well-known outer bound on the capacity region of the DMN is the {\em cut-set bound} developed by El Gamal in 1981~\cite{elgamal_81}. If the DMN can be represented by a flow network on a graph, then the cut-set bound reduces to the traditional max-flow min-cut bound which can be computed by Ford-Fulkerson algorithm~\cite{FordFulkerson1956}. The cut-set bound states that for any cut-set~$T$ of the network with nodes indexed by a set $\mathcal{I}$, the sum of the rates of transmission of messages on one side of the cut is bounded above by the conditional mutual information between the input variables in~$T$ and the output variables in~$T^c \stackrel{\mathrm{def}}{=} \mathcal{I}\setminus T$ given the input variables in~$T^c$. The DMN is a generalization of the well-studied discrete memoryless relay channel (DM-RC) \cite{CEG}. It is known that the cut-set bound is not tight (achievable) in general \cite{Aleksic}, but it is tight (achievable) for several classes of DMNs, including the degraded DM-RC~\cite{CEG}, the degraded DMN \cite{KGG05, Aref}, the semi-deterministic DM-RC~\cite{EG82}, the DM-RC with orthogonal sender components \cite{EG05}, and the linear deterministic multicast network \cite{ADT} among others.

One potential drawback of the cut-set bound is the fact that it is only a {\em weak converse} for networks with tight cut-set bound. This weak converse only guarantees that for any network with tight cut-set bound and any fixed rate vector residing outside the capacity region, the average probabilities of decoding error of any sequence of length-$n$ codes operated at the rate vector is bounded away from~$0$ as~$n$ tends to infinity. In information theory, it is also important to establish a \emph{strong converse} statement indicating that there is a sharp phase transition of the minimum achievable asymptotic error probability between rate vectors inside and outside the capacity region in the following sense: Any rate vector inside the capacity region can be supported by some sequence of length-$n$ codes with asymptotic error probability being~$0$, and the asymptotic error probability of any sequence of length-$n$ codes operated at a rate vector outside the capacity region equals~$1$.
The contrapositive of the strong converse statement can roughly be stated as follows: All codes that result in an error probability not exceeding a tolerable error $\varepsilon\in(0,1)$ as the block length grows, i.e., $\varepsilon$-reliable codes, must have rate vectors belonging to the capacity region. As a result, the strong converse establishes a sharp phase transition between achievable and non-achievable rates, ensuring that there is no tradeoff between error probability and rate as the block length approaches infinity. This motivates us to identify networks for which the strong converse property holds and to prove such strong converse statements.

\subsection{Related Work}
Behboodi and Piantanida first conjectured the strong converses for DM-RCs~\cite{Beh11} and DMNs~\cite{Beh12} with tight cut-set bound (also see the thesis by Behboodi~\cite[App.~C]{Beh_thesis}). Unfortunately it appears to the present authors that some steps of the proofs, which are based on the information spectrum method \cite{Han10}, are incomplete, as will be elaborated in the sequel after the first theorem is stated.

In our prior work \cite{FongTan16Jan}, inspired by the work of Polyanskiy and Verd\'u~\cite{PV10}, we leveraged properties of the conditional R\'enyi divergence to prove the strong converse for certain classes of DMNs with tight cut-set bound. These include the linear deterministic multicast network \cite{ADT}, the multimessage multicast networks consisting of independent DMCs \cite{networkEquivalencePartI} and the wireless erasure network \cite{dana06}, but excluding the following networks with tight cut-set bound: the degraded DM-RC~\cite{CEG}, the degraded DMN \cite{KGG05, Aref}, the semi-deterministic DM-RC~\cite{EG82}, and the DM-RC with orthogonal sender components \cite{EG05}. This work significantly strengthens our prior work by proving the strong converse for all DMNs with tight cut-set bound including the four aforementioned networks left out by our prior work. See Remark \ref{remark_compare} for a more detailed discussion.

Our generalization of the strong converse proof for DMNs to Gaussian networks is not obvious, mainly due to the fact that the strong converse proof for DMNs is based on the method of types~\cite[Ch.~2]{Csi97}. Indeed, the strong converse property does not hold for Gaussian networks with tight cut-set bound in general if long-term power constraints are used instead of almost-sure power constraints, proved by Fong and Tan for the Gaussian degraded RC~\cite{FongTan16GaussianRelay} and Truong et al.\ for the Gaussian MAC with feedback~\cite{LFT16MACfeedback}. Being aware that the existing literature lacks a conclusive statement concerning strong converses for Gaussian networks, we are motivated to provide a strong converse proof for Gaussian networks with tight cut-set bound subject to almost-sure power constraints.
\subsection{Main Contributions}
\subsubsection{First Contribution} \label{subsubSec1stContribution}
The first contribution of this work is a self-contained proof of the strong converse for DMNs with tight cut-set bound. More precisely, we prove that for a given DMN, the set of rate vectors that can be supported by a sequence of codes with asymptotic error probability equal to~$\varepsilon$ must be contained in the region prescribed by the cut-set bound as long as $\varepsilon\in [0,1)$.
The proof is based on the method of types~\cite[Ch.~2]{Csi97}. The proof techniques are inspired by the work of Csisz\'{a}r and {K\"{o}rner} \cite{CsiszarKorner82} which fully characterized the reliability function of any discrete memoryless channel (DMC) with feedback for rates above capacity and showed that feedback does not improve the reliability function. Important consequences of this result are new strong converses for the degraded DM-RC~\cite{CEG}, the general RC with feedback~\cite{CEG}, the degraded DMN \cite{KGG05, Aref}, the semi-deterministic DM-RC~\cite{EG82}, and the DM-RC with orthogonal sender components~\cite{EG05}.
\subsubsection{Second Contribution} \label{subsubSec2ndContribution}
The second contribution of this work is the generalization of our strong converse proof to Gaussian networks where the noise random variables are assumed to be additive white Gaussian and each node is subject to an almost-sure power constraint. This proof for Gaussian networks involves a careful generalization of the method of types for discrete distributions to general distributions.
More specifically, the method of types defined for DMNs is based on counting arguments since the input and output alphabets of DMNs are finite. On the contrary, the method of types defined for Gaussian networks is based on careful approximation and quantization arguments due to the continuous input and output alphabets. See Section~\ref{AWGN:subsecQuantize} for the details regarding the quantization arguments.
There is one key difference between the proof for DMNs in Section~\ref{secConverse} and the proof for Gaussian networks in Section~\ref{AWGN:secConverse}:
In the proof for Gaussian networks, we avoid using conditional types, which cannot be easily defined when the correlation between the input symbols and the noise random variables is not negligible. Instead, we modify our definition of joint type classes in Definition~\ref{AWGN:definitionJointTypeClass} so that we can omit the use of conditional types in our proof. In contrast, the proof for DMNs in Section~\ref{secConverse} relies heavily on the definition of conditional types. Important consequences of this result are new strong converses for the Gaussian degraded RC~\cite{CEG}, the general Gaussian RC with feedback~\cite{CEG}, the sender frequency-division Gaussian RC~\cite{EG05}, and the Gaussian multiple access channel (MAC) with feedback under almost-sure power constraints~\cite{ozarow84}.\footnote{Although the achievability scheme proposed by Ozarow~\cite{ozarow84} satisfies only the long-term power constraints, it can be easily modified so that the almost-sure power constraints are satisfied.}

\subsection{Paper Outline}
This paper is organized as follows. The notations used in this paper are described in the next subsection. Section~\ref{sectionDefinition} presents the problem formulation of the DMN and its capacity region for $\varepsilon\in[0,1)$, followed by the first main result in this paper --- the strong converse for DMNs with tight cut-set bound. Section~\ref{AWGN:sectionDefinition} presents the problem formulation of the Gaussian network and its capacity region for $\varepsilon\in[0,1)$, followed by the second main result of this paper --- the strong converse for Gaussian networks with tight cut-set bound. The preliminaries for the proof of the first result are contained in Section~\ref{sectionPrelim}, which includes well-known results based on the method of types. Section~\ref{secConverse} presents the proof of the first main result. The preliminaries for the proof of the second result are contained in Section~\ref{AWGN:sectionPrelim}, which explains the construction and quantization of Gaussian types. Section~\ref{AWGN:secConverse} presents the proof of the second main result. Section~\ref{conclusion} concludes this paper.
\subsection{Notation}\label{notation}
The sets of natural, real and non-negative real numbers are denoted by $\mathbb{N}$, $\mathbb{R}$ and $\mathbb{R}_+$ respectively. The $N$-dimensional identity matrix is denoted by $I_N$, the length-$N$ all-zero column vector is denoted by $0^N$, the $N_1\times N_2$ all-zero matrix is denoted by $0^{N_1\times N_2}$, and the $N_1\times N_2$ all-one matrix is denoted by $1^{N_1\times N_2}$. For any real-valued matrix $K$, we let $K^t$ denote its transpose. If $K$ is a square matrix, we let $|K|$ and $\Tr(K)$ denote the determinant and trace of $K$ respectively. If $K$ is symmetric, we use $K\succ 0$ and $K\succeq 0$ to represent that $K$ is positive definite and $K$ is positive semi-definite respectively. We let $K^{-1}$ denote the inverse of any invertible matrix~$K$. For any two real-valued matrices $A$ and $B$ of the same dimension, we use $A<B$, $A\le B$, $A\ge B$ and $A=B$ to represent the corresponding relations between $A$ and $B$ entrywise. We will take all logarithms to base~$e$ throughout this paper.

We use $\Pr\{\mathcal{E}\}$ to represent the probability of an
event~$\mathcal{E}$, and we let $\mathbf{1}\{\mathcal{E}\}$ be the indicator function of $\mathcal{E}$. Every random variable is denoted by a capital letter (e.g., $X$), and the realization and the alphabet of the random variable are denoted by the corresponding small letter (e.g., $x$) and calligraphic letter (e.g., $\mathcal{X}$) respectively.
We use $X^n$ to denote a random tuple $(X_1,  X_2,  \ldots ,  X_n)$, where the components $X_k$ have the same alphabet~$\mathcal{X}$. We let $p_X$ and $p_{Y|X}$ denote the probability distribution of $X$ and the conditional probability distribution of $Y$ given $X$ respectively for any random variables~$X$ and~$Y$ (can be both discrete, both continuous or one discrete and one continuous).
We let $p_Xp_{Y|X}$ denote the joint distribution of $(X,Y)$, i.e., $p_Xp_{Y|X}(x,y)=p_X(x)p_{Y|X}(y|x)$ for all $x$ and $y$. The expectation of $X$ is denoted by $\E[X]$.
  For any discrete random variable $(U, X,Y,Z)$ distributed according to $p_{U, X,Y,Z}$, we let $H_{p_{U,X,Y,Z}}(X|Z)$ or more simply $H_{p_{X,Z}}(X|Z)$ denote the entropy of $X$ given $Z$, and let $I_{p_{U, X,Y,Z}}(X;Y|Z)$ or more simply $I_{p_{X,Y,Z}}(X;Y|Z)$ denote the mutual information between $X$ and $Y$ given $Z$. For any $r_X$, $p_{Y|X}$ and $q_{Y|X}$ such that $r_Xp_{Y|X}$ is absolutely continuous with respect to $r_Xq_{Y|X}$, the relative entropy between $p_{Y|X}$ and $q_{Y|X}$ given $r_X$ is finite and denoted by $D(p_{Y|X}\|q_{Y|X}|r_X)$.
The $\mathcal{L}_1$-distance between two distributions $p_X$ and $q_X$ on the same discrete alphabet $\mathcal{X}$, denoted by $\|p_X-q_X\|_{\mathcal{L}_1}$, is defined as
$\|p_X-q_X\|_{\mathcal{L}_1}\stackrel{\mathrm{def}}{=}\sum_{x\in\mathcal{X}}|p_X(x)-q_X(x)|$.
For any $N$-dimensional real-valued Gaussian vector $\boldsymbol{Z}\stackrel{\mathrm{def}}{=} [Z_1\ Z_2\ \ldots Z_N]^t$ whose mean and covariance matrix are $\boldsymbol{\mu}\in \mathbb{R}^N$ and~$\mathbf{\Sigma}\in \mathbb{R}^{N\times N}$ respectively, we let
 \begin{equation}
 \mathcal{N}(\mathbf{z}; \boldsymbol{\mu}, \mathbf{\Sigma})\stackrel{\mathrm{def}}{=} \frac{1}{\sqrt{(2\pi)^N |\mathbf{\Sigma}|}}e^{-\frac{1}{2}(\mathbf{z}-\boldsymbol{\mu})^t \mathbf{\Sigma}^{-1} (\mathbf{z}-\boldsymbol{\mu}) } \label{AWGN:normalDist}
 \end{equation}
 be the corresponding probability density function.

\section{Discrete Memoryless Network and the First Main Result} \label{sectionDefinition}
We consider a general network that consists of~$N$ nodes. Let
\[
\mathcal{I}\stackrel{\mathrm{def}}{=} \{1, 2, \ldots, N\}
\]
 be the index set of the nodes.
The $N$ terminals exchange information in $n$ time slots as follows.
Node~$i$ chooses message
$W_{i,j}$ according to the uniform distribution from the alphabet
 \begin{equation}
 \mathcal{W}_{i,j}\stackrel{\mathrm{def}}{=} \{1, 2, \ldots, \lceil e^{n R_{i,j}}\rceil\} \label{defAlphabet}
 \end{equation}
 and sends $W_{i,j}$ to node~$j$ for each $(i, j)\in \mathcal{I}\times \mathcal{I}$, where $R_{i,j}$ characterizes the rate of message~$W_{i,j}$ and all the messages are mutually independent.
 For each $k\in \{1, 2, \ldots, n\}$ and each $i\in \mathcal{I}$, node~$i$ transmits $X_{i,k} \in \mathcal{X}_i$, a function of $\{W_{i,\ell} : \ell\in \mathcal{I}\}$ and $Y_i^{k-1}$, and receives $Y_{i,k} \in \mathcal{Y}_i$ in the $k^{\text{th}}$ time slot where $\mathcal{X}_i$ and $\mathcal{Y}_i$ are some alphabets that possibly depend on~$i$.
After receiving~$n$ symbols in the~$n$ time slots, node~$j$ declares~$\hat W_{i,j}$ to be the
transmitted~$W_{i,j}$ based on $\{ W_{j,\ell}: \ell \in \mathcal{I} \}$ and $Y_j^n$ for each $(i,j)\in \mathcal{I} \times \mathcal{I}$.

To simplify notation, we use the following conventions for each non-empty $T\subseteq \mathcal{I}$: For any random vector
\begin{align*}
 [X_{1} \ X_{2}\ \ldots \ X_{N}]^t &\in \mathcal{X}_1\times \mathcal{X}_2 \times \ldots \times \mathcal{X}_N,\\
\noalign {\noindent we let}
 \boldsymbol{X}&\stackrel{\mathrm{def}}{=} [X_{1} \ X_{2}\ \ldots \ X_{N}]^t\\
\noalign {\noindent be the whole vector,}
 \boldsymbol{\mathcal{X}}&\stackrel{\mathrm{def}}{=} \prod_{i=1}^N \mathcal{X}_i\\
\noalign {\noindent be the alphabet of $\boldsymbol{X}$,}
 X_T &\stackrel{\mathrm{def}}{=} [X_{i}:i\in T]^t
\end{align*}
 be the subvector of $\boldsymbol{X}$ and $\mathcal{X}_T$ be the alphabet of $X_T$.
Similarly, for any $k\in \{1, 2, \ldots, n\}$ and any random vector
\begin{align*}
  [X_{1,k} \ X_{2,k}\ \ldots \ X_{N,k}]^t &\in \mathcal{X}_1\times \mathcal{X}_2 \times \ldots \times \mathcal{X}_N,
\\
 \noalign {\noindent we let}
 \boldsymbol{X}_k&\stackrel{\mathrm{def}}{=} [X_{1,k} \ X_{2,k}\ \ldots \ X_{N,k}]^t \in \boldsymbol{\mathcal{X}}
\\
  \noalign {\noindent be the whole vector and}
X_{T,k}&\stackrel{\mathrm{def}}{=} [X_{i,k}:i\in T]^t \in \mathcal{X}_T
\end{align*}
 be the subvector of $\boldsymbol{X}_k$. For any non-empty $T_1, T_2\subseteq \mathcal{I}$ and any $N^2$-dimensional random vector
 \begin{align*}
[W_{1,1}\ W_{1,2}\ \ldots \ W_{N,N}]^t &\in \mathcal{W}_{1,1}\times \mathcal{W}_{1,2}\times \ldots \times \mathcal{W}_{N,N},
\\ \noalign {\noindent
 we let}
 \boldsymbol{W}&\stackrel{\mathrm{def}}{=}[W_{1,1}\ W_{1,2}\ \ldots \ W_{N,N}]^t
\\
   \noalign {\noindent  be the whole vector,}
 \boldsymbol{\mathcal{W}}&\stackrel{\mathrm{def}}{=}\prod_{i=1}^N\prod_{j=1}^N \mathcal{W}_{i,j}
\\ \noalign {\noindent
 be the alphabet of $\boldsymbol{W}$,}
W_{T_1\times T_2}&\stackrel{\mathrm{def}}{=}[W_{i,j}:\,(i,j)\in T_1\times T_2]
\end{align*}
 be the subvector of $ \boldsymbol{W}$, and $\mathcal{W}_{T_1\times T_2}$ be the alphabet of $W_{T_1\times T_2}$.
The following six definitions formally define a DMN and its capacity region.
\smallskip
\begin{Definition} \label{discreteMemoryless}
A discrete network consists of $N$ finite input sets
$\mathcal{X}_1, \mathcal{X}_2, \ldots, \mathcal{X}_N$, $N$ finite output sets $\mathcal{Y}_1, \mathcal{Y}_2, \ldots, \mathcal{Y}_N$ and a transition matrix $q_{\boldsymbol{Y}|\boldsymbol{X}}$. The discrete network is denoted by $(\boldsymbol{\mathcal{X}}, \boldsymbol{\mathcal{Y}}, q_{\boldsymbol{Y}|\boldsymbol{X}})$. For every non-empty $T\subsetneq\mathcal{I}$, the marginal distribution $q_{Y_{T^c}|\boldsymbol{X}}$ is defined as
\begin{equation*}
q_{Y_{T^c}|\boldsymbol{X}}(y_{T^c}|\mathbf{x})\stackrel{\mathrm{def}}{=}\sum_{y_T\in \mathcal{Y}_T}q_{\boldsymbol{Y}|\boldsymbol{X}}(\mathbf{y}|\mathbf{x}) 
\end{equation*}
for all $\mathbf{x}\in \boldsymbol{\mathcal{X}}$ and all $y_{T^c}\in \mathcal{Y}_{T^c}$.
\end{Definition}
\smallskip
\begin{Definition} \label{defCode}
An $(n, \mathbf{R})$-code, where $\mathbf{R}\stackrel{\mathrm{def}}{=}[R_{1,1}\ R_{1,2}\ \ldots\ R_{N,N}]^t \ge 0^{N^2}$ denotes the $N^2$-dimensional rate vector, for $n$ uses of the discrete network $(\boldsymbol{\mathcal{X}}, \boldsymbol{\mathcal{Y}}, q_{\boldsymbol{Y}|\boldsymbol{X}})$ consists of the
following:
\begin{enumerate}
\item A message set $\mathcal{W}_{i,j}$
 at node~$i$ for each $(i,j)\in \mathcal{I} \times \mathcal{I}$ as defined in~\eqref{defAlphabet}. Message $W_{i,j}$ is uniform on $\mathcal{W}_{i,j}$.

\item An encoding function
\[
f_{i,k} : \mathcal{W}_{\{i\}\times\mathcal{I}} \times \mathcal{Y}_i^{k-1} \rightarrow \mathcal{X}_i
 \]
 for each $i\in \mathcal{I}$ and each $k\in\{1, 2, \ldots, n\}$, where $f_{i,k}$ is the encoding function at node~$i$ in the
$k^{\text{th}}$ time slot such that\footnote{We assume by convention that the domain of $f_{i,1}$ is $\mathcal{W}_{\{i\}\times \mathcal{I}}\times \emptyset$.}
\[
X_{i,k}=f_{i,k} (W_{\{i\}\times \mathcal{I}},
Y_i^{k-1}).
\]

\item A decoding function
\[
\varphi_{i} : \mathcal{W}_{\{i\}\times\mathcal{I}} \times
\mathcal{Y}_i^{n} \rightarrow \mathcal{W}_{\mathcal{I}\times\{i\}}
 \]
 for each $i\in\mathcal{I}$, where $\varphi_{i}$ is the decoding function for $W_{\mathcal{I}\times\{i\}}$ at node~$i$ such that
 \[
 \hat W_{\mathcal{I}\times\{i\}} = \varphi_{i}(W_{\{i\}\times \mathcal{I}}, Y_i^{n}).
 \]
\end{enumerate}
\end{Definition}
\smallskip
\begin{Definition}\label{defDiscreteMemoryless}
A discrete network $(\boldsymbol{\mathcal{X}}, \boldsymbol{\mathcal{Y}}, q_{\boldsymbol{Y}|\boldsymbol{X}})$, when used multiple times, is called a \textit{discrete memoryless network (DMN)} if the following holds for any $(n, \mathbf{R})$-code:

Let $\boldsymbol{U}^{k-1}=(\boldsymbol{W}, \boldsymbol{X}^{k-1}, \boldsymbol{Y}^{k-1})$ be the collection of random variables that are generated before the $k^{\text{th}}$ time slot. Then, for each $k\in\{1, 2, \ldots, n\}$
and each $T\subsetneq\mathcal{I}$,
\begin{align*}
 \Pr\{\boldsymbol{U}^{k-1} = \mathbf{u}^{k-1}, \boldsymbol{X}_k =\mathbf{x}_k, Y_{T^c,k}=y_{T^c,k} \}
 = \Pr\{\boldsymbol{U}^{k-1} = \mathbf{u}^{k-1},  \boldsymbol{X}_k =\mathbf{x}_k \} q_{Y_{T^c}|\boldsymbol{X}}(y_{T^c,k}| \mathbf{x}_k) 
\end{align*}
holds for all $\mathbf{u}^{k-1}\in \boldsymbol{\mathcal{U}}^{k-1}$, $\mathbf{x}_k\in \boldsymbol{\mathcal{X}}$ and $y_{T^c,k}\in \mathcal{Y}_{T^c}$.
\end{Definition}
\begin{Remark}\label{remarkMemoryless}
 Definition~\ref{defDiscreteMemoryless} is consistent with the definition of a DMC with feedback stated by Massey~\cite{Massey1990}. As indicated in~\cite{Massey1990}, we cannot use $\prod_{k=1}^n q_{\boldsymbol{Y}|\boldsymbol{X}}(\mathbf{y}_k|\mathbf{x}_k)$ to define a DMN due to the presence of feedback captured by the encoding functions in Definition~\ref{defCode}.
\end{Remark}
\smallskip
\begin{Definition} \label{cutseterrorProbability}
For an $(n, \mathbf{R})$-code, we can calculate the average probability of decoding error defined as
\[
\Pr\left\{\bigcup_{i\in  \mathcal{I}} \{\varphi_{i}(W_{\{i\}\times \mathcal{I}}, Y_i^n) \ne W_{\mathcal{I}\times \{i\}}\}\right\}.
\]
We call an $(n, \mathbf{R})$-code with average probability of decoding error no larger than $\varepsilon_n$ an $(n, \mathbf{R}, \varepsilon_n)$-code.
\end{Definition}
\smallskip
\begin{Definition} \label{cutsetachievable rate}
A rate vector $\mathbf{R} \in \mathbb{R}_+^{N^2}$ is \textit{$\varepsilon$-achievable} if there exists a sequence of $(n, \mathbf{R}, \varepsilon_n)$-codes such that
$\limsup\limits_{n\rightarrow \infty}\varepsilon_n \le \varepsilon$.
\end{Definition}
\smallskip

Without loss of generality, we assume that $R_{i,i}=0$ for all $i\in \mathcal{I}$ in the rest of this paper.
\smallskip
\begin{Definition}\label{cutsetcapacity region}
The \textit{$\varepsilon$-capacity region}, denoted by $\mathcal{C}_\varepsilon$, of the DMN is the closure of the set consisting of every $\varepsilon$-achievable rate vector $\mathbf{R}$ with $R_{i,i}=0$ for all $i\in \mathcal{I}$. The \textit{capacity region} is defined to be the $0$-capacity region~$\mathcal{C}_0$.
\end{Definition}
\smallskip
The following theorem is the first main result in this paper.
\smallskip
\begin{Theorem} \label{thmMainResult}
Let $(\boldsymbol{\mathcal{X}}, \boldsymbol{\mathcal{Y}}, q_{\boldsymbol{Y}|\boldsymbol{X}})$ be a DMN. Define
\begin{equation}
\mathcal{R}_{\mathrm{cut-set}} \stackrel{\mathrm{def}}{=}   \bigcup_{p_{\boldsymbol{X}}}\bigcap_{T\subsetneq \mathcal{I}: T\ne \emptyset} \left\{ \mathbf{R}\in \mathbb{R}_+^{N^2}\left| \: \parbox[c]{2.6 in}{$ \sum\limits_{ (i,j)\in T\times T^c} R_{i,j}
 \le  I_{p_{\boldsymbol{X}}q_{Y_{T^c}|\boldsymbol{X}}}(X_T; Y_{T^c}|X_{T^c}),\\
 R_{i,i}=0 \text{ for all }i\in\mathcal{I}$} \right.\right\}. \label{Rcutset}
\end{equation}
 Then for each $\varepsilon \in [0,1)$,
\begin{equation}
\mathcal{C}_\varepsilon \subseteq \mathcal{R}_{\mathrm{cut-set}}. \label{cutsetStatement}
\end{equation}
\end{Theorem}
\smallskip
\begin{Remark}\label{remark0}
The authors in~\cite{Beh12,Beh_thesis} conjectured that the strong converse holds for general DMNs with tight cut-set bound and they employed information spectrum techniques. However, the fourth equality of the chain of equalities after equation (C.8) in \cite{Beh_thesis} need not hold, which implies that the first step of their proof in \cite[Section IV.B]{Beh12} is incomplete. Consequently, their proof has a gap. Our proof of Theorem~\ref{thmMainResult} does not use information spectrum methods. Rather, we use the method of types to establish a strong converse for DMNs with tight cut-set bound.
\end{Remark}
\smallskip
\begin{Remark} \label{remark1}
We observe from Theorem~\ref{thmMainResult} that the cut-set bound characterized by \eqref{cutsetStatement} is a universal outer bound on $\mathcal{C}_\varepsilon$ for all $0\le \varepsilon <1$, which implies the strong converse for the class of DMNs whose cut-set bounds are achievable. As mentioned in Section~\ref{subsubSec1stContribution}, the class includes the degraded DM-RC~\cite{CEG}, the general RC with feedback~\cite{CEG}, the degraded DMN~\cite{KGG05, Aref}, the semi-deterministic DM-RC~\cite{EG82}, and the DM-RC with orthogonal sender components~\cite{EG05}.
\end{Remark}
\smallskip
\begin{Remark} \label{remark_compare}
Theorem~\ref{thmMainResult} establishes the strong converse for any DMN with tight cut-set bound under the \textit{multiple unicast demand} where each node has a unique message destined for each other node.
This strong converse result strengthens our prior strong converse result~\cite{FongTan16Jan} established for some classes of DMN with tight cut-set bound under the \textit{multicast demand} where each source node sends a single message and each destination node wants to recover all the source messages. To be more explicit, our prior strong converse result specialized to the multiple unicast demand scenario states that
\begin{equation*}
\mathcal{C}_\varepsilon \subseteq {\mathcal{R}}_{\mathrm{out}}
\end{equation*}
for all $\varepsilon\in [0,1)$, where
\begin{equation}
{\mathcal{R}}_{\mathrm{out}}  \stackrel{\mathrm{def}}{=}   \bigcap_{T\subsetneq \mathcal{I}: T\ne \emptyset}\bigcup_{ p_{\boldsymbol{X}}} \left\{ \mathbf{R}\in \mathbb{R}_+^{N^2}\left| \: \parbox[c]{2.58 in}{$ \sum\limits_{ (i,j)\in T\times T^c} R_{i,j}
 \le  I_{p_{\boldsymbol{X}}q_{Y_{T^c}|\boldsymbol{X}}}(X_T; Y_{T^c}|X_{T^c}),\\
 R_{i,i}=0 \text{ for all }i\in\mathcal{I}$} \right.\right\}. \label{Rcutset2}
\end{equation}
Comparing \eqref{Rcutset} to \eqref{Rcutset2}, we observe that the union and the intersection are swapped and consequently
${\mathcal{R}}_{\mathrm{cut-set}}\subseteq {\mathcal{R}}_{\mathrm{out}}$ holds, where  the inequality is strict for many classes of networks. Thus, Theorem \ref{thmMainResult}  is considerably  stronger than the main theorem in \cite{FongTan16Jan}. In particular, Theorem \ref{thmMainResult} establishes the  strong converse for the following four networks: the physically degraded DM-RC, the physically degraded DMN, the semi-deterministic DM-RC, and the DM-RC with orthogonal sender components. Strong converses for these  important networks were not proved in our previous paper \cite{FongTan16Jan}. The proof of Theorem~\ref{thmMainResult} is based on the method of types~\cite{Csi97}, which is completely different compared to the R\'enyi divergence approach in our prior work~\cite{FongTan16Jan}. It seems challenging (to the authors) to use other standard strong converse proof techniques to prove Theorem~\ref{thmMainResult} such as the R\'enyi divergence approach~\cite{FongTan16Jan, PV10}, the information spectrum method \cite{Han10}, the blowing-up lemma \cite{Csi97,Marton86} and the reverse hypercontractivity method~\cite{LHV2017}. In particular, the hypercontractivity method seems to be most pertinent to problems whose capacity regions contain auxiliary random variables whereas the cut-set bound does not contain auxiliary random variables. Since the blowing-up lemma and the reverse hypercontractivity method are based on analyzing the product channels $\prod_{k=1}^N q_{\mathbf{Y}|\mathbf{X}}(\mathbf{y}_k|\mathbf{x}_k)$, they are not suitable for proving strong converse for a DMN with feedback due to Remark~\ref{remarkMemoryless}.
\end{Remark}

\smallskip
\begin{Remark}\label{remark3}
The proof of Theorem~\ref{thmMainResult} implies that for any fixed rate vector $\mathbf{R}$ lying outside the cut-set bound $\mathcal{R}_{\mathrm{cut-set}}$, the average probabilities of correct decoding of any sequence of $(n, \mathbf{R})$-codes tend to~0 exponentially fast. See~\eqref{convProofEq16} in the proof for the derived upper bound on the non-asymptotic probability of correct decoding. In other words, we have proved an \emph{exponential strong converse} for networks with tight cut-set bound (cf.\ Oohama's works in~\cite{oohama15} and~\cite{oohama16} that established exponential strong converse for broadcast channels). We leave the exact characterization of the strong converse exponent to future work.
\end{Remark}
\smallskip
\begin{Remark}
The proof of Theorem~\ref{thmMainResult} is inspired by two works which are based on the method of types~\cite{Csi97}. First, Tan showed in~\cite{Tan15} that the proof techniques used for analyzing the reliability functions of DMCs with feedback can be applied to DM-RCs. Second, Csisz\'ar and K\"orner~\cite{CsiszarKorner82}  fully characterized the reliability functions of any DMC with feedback for rates above capacity. We use those ideas in the proof of Theorem~\ref{thmMainResult}.
\end{Remark}
\smallskip
\begin{Example} \label{example1DM}
Consider a three-node DM-RC where the source, the relay and the destination are indexed by~1, 2 and~3 respectively. The source sends a message to the destination with the help of the relay, and we are interested in the \emph{capacity} defined as
\begin{equation}
C\stackrel{\mathrm{def}}{=}\max\left\{R_{1,3}|\,\mathbf{R}\in \mathcal{C}_0\right\}. \label{defCapacityRelayChannel}
\end{equation}
The capacity of the DM-RC is unknown in general. However, if there exists a noiseless feedback link which carries $Y_{3}^{k-1}$ to node~2 in each time slot~$k$, then the capacity of the resultant DM-RC with feedback coincides with the cut-set bound~$\max\left\{R_{1,3}|\,\mathbf{R}\in \mathcal{R}_{\mathrm{cut-set}}\right\}$ \cite[Sec.~17.4]{elgamal}, which is intuitively true because the feedback link transforms the DM-RC into a physically degraded DM-RC. Consequently, Theorem~\ref{thmMainResult} implies that the DM-RC with feedback to the relay satisfies the strong converse property. In addition, inserting two noiseless feedback links which carry $Y_{2}^{k-1}$ and $Y_{3}^{k-1}$ to node~1 in each time slot~$k$ does not further increase the capacity of the DM-RC with feedback, and hence the strong converse property also holds under this setting.
\end{Example}

\section{Gaussian Network and the Second Main Result} \label{AWGN:sectionDefinition}
In this section, we consider the Gaussian network whose channel law is described below.
 For each $k\in \{1, 2, \ldots, n\}$ and each $i\in \mathcal{I}$, node~$i$ transmits $X_{i,k} \in \mathbb{R}$, a function of $\{W_{i,\ell}:\ell\in \mathcal{I}\}$ and $Y_i^{k-1}$, and receives
 \begin{equation*}
 Y_{i,k} = \sum_{j=1}^n g_{ij}X_{j,k} + Z_{i,k}
 \end{equation*}
  in the $k^{\text{th}}$ time slot, where $g_{ij}\in \mathbb{R}$ characterizes the constant channel gain associated with the signal path starting from node~$j$ and ending at node~$i$ and $Z_{i,k}$ denotes the additive Gaussian noise experienced by node~$i$. Each node $i\in \mathcal{I}$ is subject to the almost-sure power constraint~\cite[Eq.~(17.4)]{elgamal}
  \begin{equation}
  \Pr\left\{ \frac{1}{n}\sum_{k=1}^n X_{i,k}^2 \le P_i \right\}=1 \label{AWGN:peakPowerConstraints}
  \end{equation}
where $P_i>0$ is some constant specifying the admissible power for node~$i$.
   To facilitate discussion,
  we define $\boldsymbol{X}_k\stackrel{\mathrm{def}}{=} [X_{1, k}\ X_{2, k}\ \ldots \ X_{N, k}]^t$ and $\boldsymbol{Y}_k\stackrel{\mathrm{def}}{=} [Y_{1, k}\ Y_{2, k}\ \ldots \ Y_{N, k}]^t$, and let $
\mathbf{G}\stackrel{\mathrm{def}}{=} [g_{ij}]_{(i,j)\in\mathcal{I}\times \mathcal{I}}
$
  be the $N\times N$ channel gain matrix that does not depend on~$k$. In addition, let $\boldsymbol{Z}_k\stackrel{\mathrm{def}}{=} [Z_{1, k}\ Z_{2, k}\ \ldots \ Z_{N, k}]^t$ be a zero-mean Gaussian vector with some covariance matrix~$\mathbf{\Sigma}\succ 0$ where~$\mathbf{\Sigma}$ characterizes the correlation among the~$N$ Gaussian noise random variables. The relation between $\boldsymbol{X}_k$ and $\boldsymbol{Y}_k$ can be written as follows for each $k\in\{1, 2, \ldots, n\}$:
  \begin{align}
  \boldsymbol{Y}_k = \mathbf{G} \boldsymbol{X}_k + \boldsymbol{Z}_k. \label{AWGN:inputOutputRelation}
  \end{align}
 After receiving~$n$ symbols in the~$n$ time slots, node~$j$ declares~$\hat W_{i,j}$ to be the
transmitted~$W_{i,j}$ based on $\{ W_{j,\ell}: \ell \in \mathcal{I} \}$ and $Y_j^n$ for each $(i,j)\in \mathcal{I} \times \mathcal{I}$.

To simplify notation, we use the following convention for any non-empty $T_1, T_2\subseteq \mathcal{I}$:
 For any $N\times N$ matrix
$
 \mathbf{G}= [g_{ij}]_{(i,j)\in \mathcal{I}\times \mathcal{I}}
$,
 we let
\begin{equation*}
G_{T_1\times T_2}=[g_{ij}]_{(i,j)\in T_1\times T_2}
\end{equation*}
denote the submatrix of $\mathbf{G}$.
\smallskip
\begin{Definition} \label{AWGN:defCode}
An $(n, \mathbf{R}, \mathbf{P})$-code, where $\mathbf{P}\stackrel{\mathrm{def}}{=}[P_1\ P_2\ \ldots \ P_N]^t > 0^N$ denotes the $N$-dimensional vector that specifies the admissible power, for the Gaussian network is an $(n, \mathbf{R})$-code defined in Definition~\ref{defCode} with the following extra assumptions: $\mathcal{X}=\mathcal{Y}=\mathbb{R}$ and the power constraint~\eqref{AWGN:peakPowerConstraints} is satisfied for each $i\in\mathcal{I}$.
\end{Definition}
\smallskip
\begin{Definition}\label{AWGN:defDiscreteMemoryless}
A Gaussian network, denoted by $(\mathbf{G}, \mathbf{\Sigma})$, is characterized by a channel gain matrix $\mathbf{G}\in \mathbb{R}^{N\times N}$, an $N\times N$ real-valued covariance matrix $\mathbf{\Sigma}\succ 0$, and a conditional distribution $q_{\boldsymbol{Y}|\boldsymbol{X}}$ where
\begin{equation*}
q_{\boldsymbol{Y}|\boldsymbol{X}}(\mathbf{y}|\mathbf{x})\stackrel{\mathrm{def}}{=} \mathcal{N}(\mathbf{y}; \mathbf{G} \mathbf{x}, \mathbf{\Sigma})
\end{equation*}
such that the following holds for any $(n, \mathbf{R}, \mathbf{P})$-code:

Let $\boldsymbol{U}^{k-1}=(\boldsymbol{W}, \boldsymbol{X}^{k-1}, \boldsymbol{Y}^{k-1})$ be the collection of random variables that are generated before the $k^{\text{th}}$ time slot. Then, for each $k\in\{1, 2, \ldots, n\}$
and each $T\subsetneq\mathcal{I}$,
\begin{align*}
 p_{\boldsymbol{U}^{k-1}, \boldsymbol{X}_k, Y_{T^c,k}}(\mathbf{u}^{k-1}, \mathbf{x}_k, y_{T^c,k})
 = p_{\boldsymbol{U}^{k-1}, \boldsymbol{X}_k}(\mathbf{u}^{k-1}, \mathbf{x}_k) q_{Y_{T^c}|\boldsymbol{X}}(y_{T^c,k}| \mathbf{x}_k) 
\end{align*}
holds for all $\mathbf{u}^{k-1}$, $\mathbf{x}_k$ and $y_{T^c,k}$, where $q_{Y_{T^c}|\boldsymbol{X}}$ is the marginal distribution of $q_{\boldsymbol{Y}|\boldsymbol{X}}$ defined as
\begin{equation*}
q_{Y_{T^c}|\boldsymbol{X}}(y_{T^c}|\mathbf{x})\stackrel{\mathrm{def}}{=} \begin{cases} \int_{\mathbb{R}^{|T|}}q_{\boldsymbol{Y}|\boldsymbol{X}}(\mathbf{y}|\mathbf{x}) \mathrm{d}y_T &\text{if $T\ne \emptyset$,}\\
q_{\boldsymbol{Y}|\boldsymbol{X}}(\mathbf{y}|\mathbf{x})& \text{otherwise}
\end{cases}
\end{equation*}
for all $\mathbf{x}$ and all $y_{T^c}$.
\end{Definition}
\smallskip
The average probability of decoding error for an $(n, \mathbf{R}, \mathbf{P})$-code is defined in a similar way to Definition~\ref{cutseterrorProbability}, and $\mathbf{R}\in \mathbb{R}_+^{N^2}$ is said to be \textit{$\varepsilon$-achievable} if there exists a sequence of $(n, \mathbf{R}, \mathbf{P}, \varepsilon_n)$-codes such that
$\limsup\limits_{n\rightarrow \infty}\varepsilon_n \le \varepsilon$. The \textit{$\varepsilon$-capacity region}, denoted by $\mathcal{C}_\varepsilon$, of the Gaussian network is the closure of the set consisting of every $\varepsilon$-achievable rate vector. The \textit{capacity region} is defined to be the $0$-capacity region $\mathcal{C}_0$.
The following theorem is the second main result of this paper.
\smallskip
\begin{Theorem} \label{AWGN:thmMainResult}
Let $(\mathbf{G}, \mathbf{\Sigma})$ be a Gaussian network. For each $N\times N$ covariance matrix~$\mathbf{K}$ and each non-empty $T\subsetneq \mathcal{I}$, let
$K_{T|T^c}$ denote the conditional covariance of $X_T$ given $X_{T^c}$ when  $\boldsymbol{X}\sim \mathcal{N}(\mathbf{x}; 0^N, \mathbf{K})$, i.e.,
\begin{align}
K_{T|T^c}&\stackrel{\mathrm{def}}{=}\E\Big[\E\left[(X_T -  \E[X_T|X_{T^c}]) (X_T-\E[X_T|X_{T^c}])^t\left| X_{T^c} \right.\right]\Big]. \label{AWGN:conditionalVar}
\end{align}
 Define
\begin{align}
\mathcal{S}(\mathbf{P})\stackrel{\mathrm{def}}{=}
\left\{\mathbf{K}\in \mathbb{R}^{N\times N}\left|
\text{$\mathbf{K}\succeq 0$ where the $i^{\text{th}}$ diagonal element $k_{ii}$ satisfies $k_{ii}\le P_i$ for all $i\in\mathcal{I}$}
\right.\right\} \label{AWGN:defSetS}
\end{align}
to be the set of covariance matrices that characterize the correlation among the transmitted symbols.
 Define
\begin{align}
\mathcal{R}_{\mathrm{cut-set}} \stackrel{\mathrm{def}}{=}\bigcup_{\mathbf{K}\in\mathcal{S}(\mathbf{P})}\bigcap_{T\subsetneq \mathcal{I}:T\ne \emptyset} \left\{\mathbf{R}\in \mathbb{R}_+^{N^2}\left| \: \parbox[c]{3.9 in}{$ \sum\limits_{ (i,j)\in T\times T^c} R_{i,j}
 \le  \frac{1}{2}\log\left|I_{|T^c|} + G_{T^c\times T}K_{T|T^c}G_{T^c\times T}^t \left(\Sigma_{T^c\times T^c}\right)^{-1}\right|,\\
 R_{i,i}=0 \text{ for all }i\in\mathcal{I}$} \right.\right\}. \label{AWGN:Rcutset}
\end{align}
 Then for each $\varepsilon \in [0,1)$,
\begin{equation}
\mathcal{C}_\varepsilon \subseteq \mathcal{R}_{\mathrm{cut-set}}. \label{AWGN:cutsetStatement}
\end{equation}
\end{Theorem}
\smallskip
\begin{Remark} \label{AWGN:remark1}
We observe from Theorem~\ref{AWGN:thmMainResult} that the cut-set bound characterized by \eqref{AWGN:cutsetStatement} is a universal outer bound on $\mathcal{C}_\varepsilon$ for all $0\le \varepsilon <1$, which implies the strong converse for the class of Gaussian networks whose cut-set bounds are achievable under the almost-sure power constraints~\eqref{AWGN:peakPowerConstraints}. The class includes Gaussian degraded RC~\cite{CEG}, the general Gaussian RC with feedback~\cite{CEG}, the sender frequency-division Gaussian RC~\cite{EG05}, and the Gaussian MAC with feedback.
\end{Remark}
\smallskip
\begin{Remark}\label{AWGN:remark3}
The proof of Theorem~\ref{AWGN:thmMainResult} implies that for any fixed rate vector $\mathbf{R}$ lying outside the cut-set bound $\mathcal{R}_{\mathrm{cut-set}}$, the average probabilities of correct decoding of any sequence of $(n, \mathbf{R}, \mathbf{P})$-codes tend to~0 exponentially fast. See~\eqref{AWGN:convProofEq9} in the proof for the derived upper bound on the non-asymptotic probability of correct decoding. In other words, we have proved an \emph{exponential strong converse} (cf.\ \cite{oohama15,oohama16}) for Gaussian networks with tight cut-set bound. We leave the exact characterization of the strong converse exponent to future work.
\end{Remark}
\smallskip
\begin{Example}\label{AWGN:remarkRelayChannel}
Consider a three-node Gaussian RC where the source, the relay and the destination are indexed by~1, 2 and~3 respectively. Suppose nodes~1 and~2 are subject to the almost-sure power constraints~\eqref{AWGN:peakPowerConstraints}. The capacity of this Gaussian RC as defined in~\eqref{defCapacityRelayChannel} is unknown in general. However, if there exists a noiseless feedback link which carries $Y_{3}^{k-1}$ to node~2 in each time slot~$k$, then the capacity of the resultant Gaussian RC with feedback coincides with the cut-set bound $\max\left\{R_{1,3}|\,\mathbf{R}\in \mathcal{R}_{\mathrm{cut-set}}\right\}$~\cite[Sec.~17.4]{elgamal}, which is intuitively true because the feedback link transforms the Gaussian RC into a Gaussian degraded RC. Consequently, Theorem~\ref{AWGN:thmMainResult} implies that the Gaussian RC with feedback to the relay satisfies the strong converse property. In addition, inserting two noiseless feedback links which carry $Y_{2}^{k-1}$ and $Y_{3}^{k-1}$ to node~1 in each time slot~$k$ does not further increase the capacity of the DM-RC with feedback, and hence the strong converse property also holds under this setting. However, if the almost-sure power constraints~\eqref{AWGN:peakPowerConstraints} are replaced with the long-term power constraints
 \begin{align}
\E\left[ \frac{1}{n}\sum_{k=1}^n X_{i,k}^2 \right] \le P_i,  \label{AWGN:LTPconstraints}
 \end{align}
the Gaussian RC no longer satisfies the strong converse property as proved in~\cite{FongTan16GaussianRelay}.
\end{Example}
\smallskip
\begin{Example}\label{AWGN:remarkMACfeedback}
Consider a $2$-source Gaussian MAC with feedback where the two sources are indexed by~1 and~2 respectively and the destination is indexed by~3. Suppose nodes~1 and~2 are subject to the almost-sure power constraints~\eqref{AWGN:peakPowerConstraints}. In addition, there exists a noiseless feedback link which carries $Y^{k-1}$ to both nodes~1 and~2 in each time slot~$k$. We are interested in the \emph{capacity region} defined as
\[
\mathcal{C}\stackrel{\mathrm{def}}{=}\left\{(R_{1,3},R_{2,3})|\,\mathbf{R} \in \mathcal{C}_0\right\}.
\]
Although the achievability scheme that achieves the cut-set bound $\left\{(R_{1,3}, R_{2,3})\left|\,\mathbf{R}\in\mathcal{R}_{\mathrm{cut-set}}\right.\right\}$ proposed by Ozarow~\cite{ozarow84} satisfies only the long-term power constraint in~\eqref{AWGN:LTPconstraints}, it can be easily modified so that the almost-sure power constraints~\eqref{AWGN:peakPowerConstraints} are satisfied. Therefore, the capacity region of this Gaussian MAC with feedback coincides with the cut-set bound. Consequently, Theorem~\ref{AWGN:thmMainResult} implies that the Gaussian MAC with feedback satisfies the strong converse property. However, if the almost-sure power constraints~\eqref{AWGN:peakPowerConstraints} are replaced with the long-term power constraints~\eqref{AWGN:LTPconstraints},
the Gaussian MAC with feedback no longer satisfies the strong converse property as proved in~\cite{LFT16MACfeedback}.
\end{Example}

\section{Preliminaries for Proving Theorem~\ref{thmMainResult} --- The Method of Types} \label{sectionPrelim}
The following definitions and results are standard \cite[Ch.~2]{Csi97}.
The \textit{type} of a sequence $x^n \in \mathcal{X}^n$, denoted by $\phi_X^{[x^n]}$, is the empirical distribution of $x^n$, i.e.,
\begin{equation*}
\phi_X^{[x^n]}(a) \stackrel{\mathrm{def}}{=} \frac{N(a| x^n)}{n}
\end{equation*}
for all $a\in \mathcal{X}$ where $N(a| x^n)$ denotes the number of occurrences of the symbol $a$ in $x^n$. The set of all possible types of sequences in $\mathcal{X}^n$ is denoted by
\begin{equation*}
\mathcal{P}_n(\mathcal{X}) \stackrel{\mathrm{def}}{=} \left\{ \left.\phi_X^{[x^n]} \, \right| x^n \in \mathcal{X}^n  \right\}.
\end{equation*}
Similarly, the set of all possible types of sequences in $\mathcal{Y}^n$ conditioned on a type $r_X\in \mathcal{P}_n(\mathcal{X})$ is denoted by
\begin{equation*}
\mathcal{P}_n(\mathcal{Y}|r_X) \stackrel{\mathrm{def}}{=} \left\{ \left.s_{Y|X} \, \right| \text{There exists an $(x^n, y^n)$ such that $\phi_X^{[x^n]}=r_X$ and $\phi_{X,Y}^{[(x^n, y^n)]}=r_X s_{Y|X}$}  \right\}.
\end{equation*}
For a given type $r_X \in \mathcal{P}_n(\mathcal{X})$, the \textit{type class} of $r_X$ is defined as
\begin{equation*}
\mathcal{T}_{r_X}^{(n)} \stackrel{\mathrm{def}}{=} \left\{x^n\in \mathcal{X}^n \left|\, \phi_{X}^{[x^n]}  = r_X \right.\right\}.
\end{equation*}
A well-known upper bound on the number of types is
\begin{equation}
\left|\mathcal{P}_n(\mathcal{X}) \right| \le (n+1)^{|\mathcal{X}|}. \label{typeClassBound}
\end{equation}
We will frequently use the following fact without explicit explanation: For each $r_X \in \mathcal{P}_n(\mathcal{X})$, each $s_{Y|X}\in \mathcal{P}_n(\mathcal{Y}|r_X)$ and each transition matrix $q_{Y|X}$, the following equality holds for any $(x^n,y^n) \in \mathcal{T}_{r_X s_{Y|X}}^{(n)}$:
\begin{align*}
\prod_{k=1}^n q_{Y|X}(y_k|x_k)&=\prod_{x, y} q_{Y|X}(y|x)^{n r_X(x)s_{Y|X}(y|x)}\\
&= e^{-n(H_{r_Xs_{Y|X}}(Y|X)+D(s_{Y|X}\|q_{Y|X} |r_X))}.
\end{align*}

\section{Proof of Theorem~\ref{thmMainResult}}\label{secConverse}
In this section, we will show that
\begin{equation}
\mathcal{C}_\varepsilon \subseteq  \mathcal{R}_{\mathrm{cut-set}} \label{convProofSt}
\end{equation}
 for all $\varepsilon\in[0,1)$ where $\mathcal{R}_{\mathrm{cut-set}}$ is as defined in~\eqref{Rcutset}. It suffices to show that for any $\mathbf{R} \notin \mathcal{R}_{\mathrm{cut-set}}$ and any sequence of $(n, \mathbf{R}, \varepsilon_n)$-codes,
 \begin{equation}
\lim_{n\rightarrow \infty}\varepsilon_n =1. \label{convProofError}
 \end{equation}
 To this end, we fix a rate vector~$\mathbf{R} \notin \mathcal{R}_{\mathrm{cut-set}}$ and a sequence of $(n, \mathbf{R} , \varepsilon_n)$-codes.
  \subsection{Relating $\mathbf{R}$ to the Cut-Set Bound}
  Since $\mathbf{R} \notin \mathcal{R}_{\mathrm{cut-set}}$ and $\mathcal{R}_{\mathrm{cut-set}}$ is closed, we can always find a positive number denoted by $\delta>0$ such that for any distribution $r_{\boldsymbol{X}}$ defined on $\boldsymbol{\mathcal{X}}$, there exists a non-empty $V_{r_{\boldsymbol{X}}}\subsetneq\mathcal{I}$ that satisfies
\begin{equation}
\sum_{(i,j)\in V_r\times V_r^c}R_{i,j}\ge  I_{r_{\boldsymbol{X}}q_{Y_{V_r^c}|\boldsymbol{X}}}(X_{V_r}; Y_{V_r^c}|X_{V_r^c})+\delta, \label{convProofRate}
 \end{equation}
 where the shorthand notation $V_r$ is used to denote $V_{r_{\boldsymbol{X}}}$.
\subsection{Simplifying the Correct Decoding Probability by Using the Discrete Memoryless Property}
Fix a natural number~$n$ and let $p_{\boldsymbol{W} , \boldsymbol{X}^n, \boldsymbol{Y}^n, \hat{\boldsymbol{W}} }$ be the probability distribution induced by the $(n, \mathbf{R}, \varepsilon_n)$-code. Unless specified otherwise, the probabilities are evaluated according to~$p_{\boldsymbol{W} , \boldsymbol{X}^n, \boldsymbol{Y}^n, \hat{\boldsymbol{W}} }$ in the rest of the proof.
Consider the probability of correct decoding
\begin{align}
1-\varepsilon_n = \frac{1}{|\boldsymbol{\mathcal{W}}|}\sum_{\mathbf{w}\in \boldsymbol{\mathcal{W}}}\Pr\left\{\left. \bigcap_{i\in \mathcal{I}} \left\{\varphi_i\left(w_{\{i\}\times\mathcal{I}},Y_i^n\right) = w_{\mathcal{I}\times\{i\}}\right\}\right|\boldsymbol{W}=\mathbf{w}\right\}. \label{convProofEq1}
\end{align}
In order to simplify the right-hand side (RHS) of~\eqref{convProofEq1}, we write for each $\mathbf{w}\in \boldsymbol{\mathcal{W}}$
\begin{align}
&\Pr\left\{\left. \bigcap_{i\in \mathcal{I}} \left\{\varphi_i(w_{\{i\}\times \mathcal{I}},Y_i^n) = w_{\mathcal{I}\times\{i\}}\right\}\right|\boldsymbol{W}=\mathbf{w}\right\} \notag\\
& \quad= \sum_{\mathbf{y}^n \in \boldsymbol{\mathcal{Y}}^n} p_{\boldsymbol{Y}^n|\boldsymbol{W}=\mathbf{w}}(\mathbf{y}^n) \times \mathbf{1}\left\{\bigcap_{i\in \mathcal{I}} \left\{\varphi_i(w_{\{i\}\times \mathcal{I}},y_i^n) = w_{\mathcal{I}\times\{i\}}\right\}\right\}\notag\\
&\quad = \sum_{\mathbf{y}^n\in \boldsymbol{\mathcal{Y}}^n} \prod_{k=1}^n  p_{\boldsymbol{Y}_k|\boldsymbol{Y}^{k-1}, W_{\{i\}\times \mathcal{I}}=w_{\{i\}\times \mathcal{I}}}(\mathbf{y}_k|\mathbf{y}^{k-1})\times \mathbf{1}\left\{\bigcap_{i\in \mathcal{I}} \left\{\varphi_i(w_{\{i\}\times \mathcal{I}},y_i^n) = w_{\mathcal{I}\times\{i\}}\right\}\right\}\notag\\
&\quad \stackrel{\text{(a)}}{=} \sum_{\mathbf{y}^n\in \boldsymbol{\mathcal{Y}}^n} \prod_{k=1}^n  p_{\boldsymbol{Y}_k|\boldsymbol{X}_k}(\mathbf{y}_k|(f_{i,k}(w_{\{i\}\times \mathcal{I}}, y_i^{k-1}):i\in\mathcal{I}))\times \mathbf{1}\left\{\bigcap_{i\in \mathcal{I}} \left\{\varphi_i(w_{\{i\}\times \mathcal{I}},y_i^n) = w_{\mathcal{I}\times\{i\}}\right\}\right\}, \label{convProofEq2Temp}
\end{align}
where (a) follows from the equality below for each $\mathbf{y}^n\in \boldsymbol{\mathcal{Y}}^n$ and each $k\in\{1, 2, \ldots, n\}$ due to the definition of the code in Definition~\ref{defCode} and the discrete memoryless property of the network implied by Definition~\ref{defDiscreteMemoryless}:
\begin{align*}
p_{\boldsymbol{Y}_k|\boldsymbol{Y}^{k-1}, W_{\{i\}\times \mathcal{I}}=w_{\{i\}\times \mathcal{I}}}(\mathbf{y}_k|\mathbf{y}^{k-1}) &= p_{\boldsymbol{Y}_k|\boldsymbol{X}_k, \boldsymbol{Y}^{k-1}, W_{\{i\}\times \mathcal{I}}=w_{\{i\}\times \mathcal{I}}}(\mathbf{y}_k|(f_{i,k}(w_{\{i\}\times \mathcal{I}}, y_i^{k-1}):i\in\mathcal{I}),\mathbf{y}^{k-1}) \\
& = p_{\boldsymbol{Y}_k|\boldsymbol{X}_k}(\mathbf{y}_k|(f_{i,k}(w_{\{i\}\times \mathcal{I}}, y_i^{k-1}):i\in\mathcal{I})).
\end{align*}
 In order to simplify notation, we define the following for every $T\subsetneq \mathcal{I}$: $\hat w_{\mathcal{I}\times \{i\}} \stackrel{\mathrm{def}}{=} \varphi_i(w_{\{i\}\times \mathcal{I}}, y_i^n)$,
$
\hat w_{\mathcal{I}\times T^c} \stackrel{\mathrm{def}}{=} (\hat w_{\mathcal{I}\times \{i\}} : i\in T^c)$, $\hat{\mathbf{w}} \stackrel{\mathrm{def}}{=} \hat w_{\mathcal{I}\times \mathcal{I}}$,
\begin{align*}
x_{i,k}(w_{\{i\}\times \mathcal{I}}, y_i^{k-1})&\stackrel{\mathrm{def}}{=}f_{i,k}(w_{\{i\}\times \mathcal{I}}, y_i^{k-1}),\\
x_{{T^c},k}(w_{{T^c}\times \mathcal{I}}, y_{T^c}^{k-1})&\stackrel{\mathrm{def}}{=} (x_{i,k}(w_{\{i\}\times \mathcal{I}}, y_i^{k-1}): i\in {T^c}),\\
\mathbf{x}_k(\mathbf{w}, \mathbf{y}^{k-1})&\stackrel{\mathrm{def}}{=} x_{\mathcal{I},k}(w_{{\mathcal{I}}\times \mathcal{I}}, y_{\mathcal{I}}^{k-1}),\\
x_{T^c}^n(w_{{T^c}\times \mathcal{I}}, y_{T^c}^{n-1}) &\stackrel{\mathrm{def}}{=} (x_{{T^c},1}(w_{{T^c}\times \mathcal{I}}), x_{{T^c},2}(w_{{T^c}\times \mathcal{I}}, y_{{T^c},1}), \ldots, x_{{T^c},n}(w_{{T^c}\times \mathcal{I}}, y_{T^c}^{n-1}))\\
\noalign{\noindent and}
\mathbf{x}^n(\mathbf{w}, \mathbf{y}^{n-1}) &\stackrel{\mathrm{def}}{=} (\mathbf{x}_1(\mathbf{w}), \mathbf{x}_2(\mathbf{w}, \mathbf{y}_{1}), \ldots, \mathbf{x}_n(\mathbf{w}, \mathbf{y}^{n-1})).
\end{align*}
Rewrite~\eqref{convProofEq2Temp} as
\begin{align}
\Pr\left\{\left. \bigcap_{i\in \mathcal{I}} \left\{\varphi_i(w_{\{i\}\times \mathcal{I}},Y_i^n) = w_{\mathcal{I}\times\{i\}}\right\}\right|\boldsymbol{W}=\mathbf{w}\right\} = \sum_{\mathbf{y}^n\in \boldsymbol{\mathcal{Y}}^n} \prod_{k=1}^n  p_{\boldsymbol{Y}_k|\boldsymbol{X}_k}(\mathbf{y}_k|\mathbf{x}_k(\mathbf{w}, \mathbf{y}^{k-1})) \times \mathbf{1}\left\{\hat{\mathbf{w}}=\mathbf{w}\right\}. \label{convProofEq2}
\end{align}

\subsection{Further Simplifying the Correct Decoding Probability by Using the Method of Types}
For each $\mathbf{w}\in\boldsymbol{\mathcal{W}}$, each type
$
r_{\boldsymbol{X}}\in \mathcal{P}_n(\boldsymbol{\mathcal{X}})$
 and each conditional type
$
 s_{\boldsymbol{Y}|\boldsymbol{X}}\in \mathcal{P}_n(\boldsymbol{\mathcal{Y}}|r_{\boldsymbol{X}})$,
we define
\begin{equation}
\mathcal{A}(\mathbf{w}; r_{\boldsymbol{X}}, s_{\boldsymbol{Y}|\boldsymbol{X}})\stackrel{\mathrm{def}}{=} \left\{\mathbf{y}^n \in \boldsymbol{\mathcal{Y}}^n\left|\left(\mathbf{x}^n(\mathbf{w},\mathbf{y}^{n-1}), \mathbf{y}^n\right)\in \mathcal{T}_{r_{\boldsymbol{X}}s_{\boldsymbol{Y}|\boldsymbol{X}}}^{(n)}\right.\right\} \label{defSetA}
\end{equation}
and define for each non-empty $T\subsetneq\mathcal{I}$ and each $w_{T^c\times \mathcal{I}}\in \mathcal{W}_{T^c\times \mathcal{I}}$
\begin{equation}
\mathcal{F}_T(w_{T^c\times \mathcal{I}};r_{\boldsymbol{X}}, s_{Y_{T^c}|\boldsymbol{X}})\stackrel{\mathrm{def}}{=} \left\{
y_{T^c}^n\in \mathcal{Y}_{T^c}^n\left|\,\parbox[c]{3.5 in}{$(x_{T^c}^n(w_{T^c\times\mathcal{I}}, y_{T^c}^n), y_{T^c}^n)\in \mathcal{T}_{u_{X_{T^c}, Y_{T^c}}}^{(n)}$ where
$u_{X_{T^c}, Y_{T^c}}$ is the marginal type of $r_{\boldsymbol{X}}s_{Y_{T^c}|\boldsymbol{X}}$ restricted to $(X_{T^c}, Y_{T^c})$}
\right.\right\}. \label{defSetF}
\end{equation}
Note that the set $\mathcal{A}(\mathbf{w}; r_{\boldsymbol{X}}, s_{\boldsymbol{Y}|\boldsymbol{X}})$ in~\eqref{defSetA} also plays a crucial role in the proof of the upper bound on the reliability functions for DM-RCs in~\cite{Tan15}.
Following~\eqref{convProofEq2} and adopting the shorthand notation 
$\mathcal{A}(\mathbf{w}; r, s)$ to denote the set in~\eqref{defSetA}, since the sets in the collection $\left\{\left.\mathcal{A}(\mathbf{w}; r, s)\right|r_{\boldsymbol{X}}\in \mathcal{P}_n(\boldsymbol{\mathcal{X}}),
 s_{\boldsymbol{Y}|\boldsymbol{X}}\in \mathcal{P}_n(\boldsymbol{\mathcal{Y}}|r_{\boldsymbol{X}})\right\}$ form a partition on $\boldsymbol{\mathcal{Y}}^n$ such that
  \[
  \boldsymbol{\mathcal{Y}}^n = \bigcup_{r_{\boldsymbol{X}}\in\mathcal{P}_n(\boldsymbol{\mathcal{X}})} \bigcup_{ s_{\boldsymbol{Y}|\boldsymbol{X}}\in \mathcal{P}_n(\boldsymbol{\mathcal{Y}}|r_{\boldsymbol{X}})} \mathcal{A}(\mathbf{w}; r_{\boldsymbol{X}}, s_{\boldsymbol{Y}|\boldsymbol{X}})
  \]
  and
  \[
   \mathcal{A}(\mathbf{w}; r_{\boldsymbol{X}}, s_{\boldsymbol{Y}|\boldsymbol{X}}) \cap  \mathcal{A}(\mathbf{w}; r_{\boldsymbol{X}}^\prime, s_{\boldsymbol{Y}|\boldsymbol{X}}^\prime) = \emptyset
  \]
  for any $( r_{\boldsymbol{X}},  s_{\boldsymbol{Y}|\boldsymbol{X}}) \ne (r_{\boldsymbol{X}}^\prime,s_{\boldsymbol{Y}|\boldsymbol{X}}^\prime)$, we have
\begin{align}
 & \sum_{\mathbf{y}^n\in \boldsymbol{\mathcal{Y}}^n} \prod_{k=1}^n  p_{\boldsymbol{Y}_k|\boldsymbol{X}_k}(\mathbf{y}_k|\mathbf{x}_k(\mathbf{w}, \mathbf{y}^{k-1})) \times \mathbf{1}\left\{\hat{\mathbf{w}}=\mathbf{w}\right\} \notag\\
 &\quad = \sum_{r_{\boldsymbol{X}}\in\mathcal{P}_n(\boldsymbol{\mathcal{X}})}\sum_{s_{\boldsymbol{Y}|\boldsymbol{X}}\in \mathcal{P}_n(\boldsymbol{\mathcal{Y}}|r_{\boldsymbol{X}})}\sum\limits_{\substack{\mathbf{y}^n \in\\ \mathcal{A}(\mathbf{w}; r, s)}}\prod_{k=1}^n  p_{\boldsymbol{Y}_k|\boldsymbol{X}_k}(\mathbf{y}_k|\mathbf{x}_k(\mathbf{w}, \mathbf{y}^{k-1})) \times\mathbf{1}\left\{\hat{\mathbf{w}}=\mathbf{w}\right\}. \label{convProofEq3}
 \end{align}
  \subsection{Bounding the Correct Decoding Probability in Terms of $\mathcal{F}_T(w_{T^c\times \mathcal{I}};r,s)$}
Fix any arbitrary non-empty $T\subsetneq\mathcal{I}$. Define
  \begin{equation}
  a_T(r,s)\stackrel{\mathrm{def}}{=}  H_{r_{\boldsymbol{X}}s_{Y_{T^c}|\boldsymbol{X}}}(Y_{T^c}|\boldsymbol{X})+D(s_{Y_{T^c}|\boldsymbol{X}}\|q_{ Y_{T^c}|\boldsymbol{X}} |r_{\boldsymbol{X}}) \label{defa}
  \end{equation}
  to simplify notation. In order to simplify the RHS of~\eqref{convProofEq3}, we consider the innermost product therein. In particular, we consider the following chain of equalities for each $r_{\boldsymbol{X}}\in\mathcal{P}_n(\boldsymbol{\mathcal{X}})$, each $s_{\boldsymbol{Y}|\boldsymbol{X}}\in \mathcal{P}_n(\boldsymbol{\mathcal{Y}}|r_{\boldsymbol{X}})$, each $\mathbf{w}\in \boldsymbol{\mathcal{W}}$, and each $\mathbf{y}^n \in \mathcal{A}(\mathbf{w}; r, s)$:
  \begin{align}
 & \prod_{k=1}^n  p_{\boldsymbol{Y}_k|\boldsymbol{X}_k}(\mathbf{y}_k|\mathbf{x}_k(\mathbf{w}, \mathbf{y}^{k-1})) \notag\\
 &\quad = \prod_{k=1}^n  p_{Y_{T^c,k}|\boldsymbol{X}_k}(y_{T^c,k}|\mathbf{x}_k(\mathbf{w}, \mathbf{y}^{k-1}))  p_{Y_{T,k}|\boldsymbol{X}_k,Y_{T^c,k}}(y_{T,k}|\mathbf{x}_k(\mathbf{w}, \mathbf{y}^{k-1}), y_{T^c, k})\notag\\
 &\quad \stackrel{\text{(b)}}{=} \Bigg(\prod_{\substack{\mathbf{x}, y_{T^c}}}  q_{Y_{T^c}|\boldsymbol{X}}(y_{T^c}|\mathbf{x})^{nr(\mathbf{x})s(y_{T^c}|\mathbf{x})}\Bigg)\Bigg( \prod_{k=1}^n p_{Y_{T,k}|\boldsymbol{X}_k,Y_{T^c,k}}(y_{T,k}|\mathbf{x}_k(\mathbf{w}, \mathbf{y}^{k-1}), y_{T^c, k})\Bigg) \notag\\
 &\quad\stackrel{\eqref{defa}}{=} e^{-n a_T(r,s)} \prod_{k=1}^n p_{Y_{T,k}|\boldsymbol{X}_k,Y_{T^c,k}}(y_{T,k}|\mathbf{x}_k(\mathbf{w}, \mathbf{y}^{k-1}), y_{T^c, k}) \label{convProofEq3*}
  \end{align}
  where (b) follows from Definition~\ref{defDiscreteMemoryless} and the fact that $\mathbf{y}^n \in \mathcal{A}(\mathbf{w}; r, s)$ (recall the definition of $\mathcal{A}(\mathbf{w}; r, s)$ in \eqref{defSetA}).
 Following~\eqref{convProofEq3} and letting $\mathcal{F}_T(w_{T^c\times \mathcal{I}};r,s)$ denote the set in~\eqref{defSetF}, we consider the following chain of inequalities for each $r_{\boldsymbol{X}}\in\mathcal{P}_n(\boldsymbol{\mathcal{X}})$ and each $s_{\boldsymbol{Y}|\boldsymbol{X}}\in \mathcal{P}_n(\boldsymbol{\mathcal{Y}}|r_{\boldsymbol{X}})$:
 \begin{align}
& \sum_{\mathbf{w}\in \boldsymbol{\mathcal{W}}}\sum\limits_{\substack{\mathbf{y}^n \in\\ \mathcal{A}(\mathbf{w}; r, s)}}\prod_{k=1}^n  p_{\boldsymbol{Y}_k|\boldsymbol{X}_k}(\mathbf{y}_k|\mathbf{x}_k(\mathbf{w}, \mathbf{y}^{k-1})) \times \mathbf{1}\left\{\hat{\mathbf{w}}=\mathbf{w}\right\} \notag\\
&\quad \stackrel{\eqref{convProofEq3*}}{=} e^{-n a_T(r,s)} \sum_{\mathbf{w}\in \boldsymbol{\mathcal{W}}}\sum\limits_{\substack{\mathbf{y}^n \in\\ \mathcal{A}(\mathbf{w}; r, s)}}          \prod_{k=1}^n p_{Y_{T,k}|\boldsymbol{X}_k,Y_{T^c,k}}(y_{T,k}|\mathbf{x}_k(\mathbf{w}, \mathbf{y}^{k-1}), y_{T^c, k})  \times \mathbf{1}\left\{\hat{\mathbf{w}}=\mathbf{w}\right\}\notag \\
         &\quad \stackrel{\text{(c)}}{\le} e^{-n a_T(r,s)} \sum_{\mathbf{w}\in \boldsymbol{\mathcal{W}}}\sum\limits_{\substack{y_{T^c}^n \in \\ \mathcal{F}_T(w_{T^c\times \mathcal{I}};r,s)}} \sum_{y_T^n\in\mathcal{Y}_T^n}  \prod_{k=1}^n p_{Y_{T,k}|\boldsymbol{X}_k,Y_{T^c,k}}(y_{T,k}|\mathbf{x}_k(\mathbf{w}, \mathbf{y}^{k-1}), y_{T^c, k}) \times \mathbf{1}\left\{ \hat w_{\mathcal{I}\times T^c} = w_{\mathcal{I}\times T^c}\right\}\notag\\
 &\quad \stackrel{\text{(d)}}{=} e^{-n a_T(r,s)} \sum_{\mathbf{w}\in \boldsymbol{\mathcal{W}}}\sum\limits_{\substack{y_{T^c}^n \in \mathcal{F}_T(w_{T^c\times \mathcal{I}};r,s)}} \mathbf{1}\left\{ \hat w_{\mathcal{I}\times T^c} = w_{\mathcal{I}\times T^c}\right\} \notag\\
 &\quad = e^{-n a_T(r,s)} \sum_{w_{(T\times T^c)^c}\in \mathcal{W}_{(T\times T^c)^c}}\sum\limits_{\substack{y_{T^c}^n \in \mathcal{F}_T(w_{T^c\times \mathcal{I}};r,s)}}  \sum_{w_{T\times T^c}\in \mathcal{W}_{T\times T^c}}\mathbf{1}\left\{\hat w_{\mathcal{I}\times T^c} = w_{\mathcal{I}\times T^c}\right\}\notag\\
 &\quad \stackrel{\text{(e)}}{\le} e^{-n a_T(r,s)}  \sum_{w_{(T\times T^c)^c}\in \mathcal{W}_{(T\times T^c)^c}}\sum\limits_{\substack{y_{T^c}^n \in \mathcal{F}_T(w_{T^c\times \mathcal{I}};r,s)}} 1  \notag\\
 & \quad= e^{-n a_T(r,s)}\sum_{w_{(T\times T^c)^c}\in \mathcal{W}_{(T\times T^c)^c}}|\mathcal{F}_T(w_{T^c\times \mathcal{I}};r,s)|, \label{convProofEq4}
\end{align}
where
\begin{enumerate}
\item[(c)] follows from the definitions of $\mathcal{A}(\mathbf{w}; r, s)$ and $\mathcal{F}_T(w_{T^c\times \mathcal{I}};r,s)$ in~\eqref{defSetA} and~\eqref{defSetF} respectively.
\item[(d)] follows from the fact that $\mathbf{1}\left\{ \hat w_{\mathcal{I}\times T^c} = w_{\mathcal{I}\times T^c}\right\}$ is a function of $(\mathbf{w}, y_{T^c}^n)$
\item[(e)] follows from the inequality below which is due to the fact that $\hat w_{\mathcal{I}\times T^c}$ is a function of $(w_{(T\times T^c)^c},y_{T^c}^n)$:
   \begin{align*}
    &\sum_{w_{T\times T^c}\in \mathcal{W}_{T\times T^c}}\mathbf{1}\left\{\hat w_{\mathcal{I}\times T^c} = w_{\mathcal{I}\times T^c}\right\} \le 1
    \end{align*}
   for each  $(w_{(T\times T^c)^c},y_{T^c}^n)\in \mathcal{W}_{(T\times T^c)^c}\times \mathcal{Y}_{T^c}^n$.
    \end{enumerate}
    \subsection{Bounding the Size of $\mathcal{F}_T(w_{T^c\times \mathcal{I}};r,s)$}
For each $r_{\boldsymbol{X}}\in\mathcal{P}_n(\boldsymbol{\mathcal{X}})$ and each $s_{\boldsymbol{Y}|\boldsymbol{X}}\in \mathcal{P}_n(\boldsymbol{\mathcal{Y}}|r_{\boldsymbol{X}})$, we let $u_{X_{T^c}, Y_{T^c}}$ denote the marginal type induced by $r_{\boldsymbol{X}}s_{Y_{T^c}|\boldsymbol{X}}$ in order to obtain an upper bound on $|\mathcal{F}_T(w_{T^c\times \mathcal{I}};r,s)|$ as follows. For each $r_{\boldsymbol{X}}\in\mathcal{P}_n(\boldsymbol{\mathcal{X}})$, each $s_{\boldsymbol{Y}|\boldsymbol{X}}\in \mathcal{P}_n(\boldsymbol{\mathcal{Y}}|r_{\boldsymbol{X}})$ and each $w_{T^c\times \mathcal{I}}\in \mathcal{W}_{T^c\times \mathcal{I}}$, since
\begin{align*}
\sum_{y_{T^c}^n\in \mathcal{F}_T(w_{T^c\times \mathcal{I}};r,s)} \prod_{k=1}^n u_{Y_{T^c}|X_{T^c}}(y_{T^c,k}|x_{T^c, k}(w_{T^c\times \mathcal{I}}, y_{T^c}^{k-1})) \le 1,
\end{align*}
it follows that
\begin{align*}
\sum_{y_{T^c}^n\in \mathcal{F}_T(w_{T^c\times \mathcal{I}};r,s)} \prod_{x_{T^c}, y_{T^c}} u_{Y_{T^c}|X_{T^c}}(y_{T^c}|x_{T^c})^{nu_{X_{T^c}, Y_{T^c}}(x_{T^c}, y_{T^c})} \le 1
\end{align*}
(recall the definition of~$\mathcal{F}_T(w_{T^c\times \mathcal{I}};r,s)$ in~\eqref{defSetF}),
which implies that
\begin{align*}
\sum_{y_{T^c}^n\in \mathcal{F}_T(w_{T^c\times \mathcal{I}};r,s)} e^{-nH_{u_{X_{T^c}, Y_{T^c}}}(Y_{T^c}|X_{T^c})} \le 1,
\end{align*}
which then implies that
\begin{align}
|\mathcal{F}_T(w_{T^c\times \mathcal{I}};r,s)| &\le e^{nH_{u_{X_{T^c}, Y_{T^c}}}(Y_{T^c}|X_{T^c})} \notag \\
& = e^{nH_{r_{\boldsymbol{X}}s_{Y_{T^c}|\boldsymbol{X}}}(Y_{T^c}|X_{T^c})}. \label{convProofEq5}
\end{align}
Combining~\eqref{convProofEq4}, \eqref{defa} and~\eqref{convProofEq5} and using the fact due to \eqref{defa} that
\begin{equation*}
\frac{|\mathcal{W}_{(T\times T^c)^c}|}{|\boldsymbol{\mathcal{W}}|} = \frac{1}{\prod\limits_{(i,j)\in T\times T^c}\lceil e^{n R_{i,j}}\rceil} \le e^{-n\sum\limits_{(i,j)\in T\times T^c}R_{i,j}},
\end{equation*}
 we have
for each $r_{\boldsymbol{X}}\in\mathcal{P}_n(\boldsymbol{\mathcal{X}})$ and each $s_{\boldsymbol{Y}|\boldsymbol{X}}\in \mathcal{P}_n(\boldsymbol{\mathcal{Y}}|r_{\boldsymbol{X}})$
\begin{align}
 &\frac{1}{|\boldsymbol{\mathcal{W}}|}\sum_{\mathbf{w}\in \boldsymbol{\mathcal{W}}}\sum\limits_{\substack{\mathbf{y}^n \in\\ \mathcal{A}(\mathbf{w}; r, s)}}\prod_{k=1}^n  p_{\boldsymbol{Y}_k|\boldsymbol{X}_k}(\mathbf{y}_k|\mathbf{x}_k(\mathbf{w}, \mathbf{y}^{k-1})) \times \mathbf{1}\left\{\hat{\mathbf{w}}=\mathbf{w}\right\} \notag\\
 &\quad \le e^{-n\big(\sum\limits_{(i,j)\in T\times T^c}R_{i,j}-I_{r_{\boldsymbol{X}}s_{Y_{T^c}|\boldsymbol{X}}}(X_T; Y_{T^c}|X_{T^c})+D(s_{Y_{T^c}|\boldsymbol{X}}\|q_{Y_{T^c}|\boldsymbol{X}} |r_{\boldsymbol{X}})\big)}. \label{convProofEq6}
\end{align}
Note that~\eqref{convProofEq6} resembles \cite[Eq.~(5)]{CsiszarKorner82} in the proof of the reliability functions for DMCs with feedback.
  \subsection{Bounding the Correct Decoding Probability in Terms of $\mathcal{A}(\mathbf{w}; r, s)$}
 We now bound the left-hand side (LHS) of~\eqref{convProofEq6} in another way for each $r_{\boldsymbol{X}}\in\mathcal{P}_n(\boldsymbol{\mathcal{X}})$ and each $s_{\boldsymbol{Y}|\boldsymbol{X}}\in \mathcal{P}_n(\boldsymbol{\mathcal{Y}}|r_{\boldsymbol{X}})$ as follows:
\begin{align}
 &\frac{1}{|\boldsymbol{\mathcal{W}}|}\sum_{\mathbf{w}\in \boldsymbol{\mathcal{W}}}\sum\limits_{\substack{\mathbf{y}^n \in\\ \mathcal{A}(\mathbf{w}; r, s)}}\prod_{k=1}^n  p_{\boldsymbol{Y}_k|\boldsymbol{X}_k}(\mathbf{y}_k|\mathbf{x}_k(\mathbf{w}, \mathbf{y}^{k-1}))  \times \mathbf{1}\left\{\hat{\mathbf{w}}=\mathbf{w}\right\} \notag\\
 & \quad\le \frac{1}{|\boldsymbol{\mathcal{W}}|}\sum_{\mathbf{w}\in \boldsymbol{\mathcal{W}}}\sum\limits_{\substack{\mathbf{y}^n \in\\ \mathcal{A}(\mathbf{w}; r, s)}}\prod_{k=1}^n  p_{\boldsymbol{Y}_k|\boldsymbol{X}_k}(\mathbf{y}_k|\mathbf{x}_k(\mathbf{w}, \mathbf{y}^{k-1})) \notag\\
 & \quad \stackrel{\text{(f)}}{=}\frac{1}{|\boldsymbol{\mathcal{W}}|}\sum_{\mathbf{w}\in \boldsymbol{\mathcal{W}}}\sum\limits_{\substack{\mathbf{y}^n \in\\ \mathcal{A}(\mathbf{w}; r, s)}} \prod_{\substack{\mathbf{x}, \mathbf{y}}}  q_{\boldsymbol{Y}|\boldsymbol{X}}(\mathbf{y}|\mathbf{x})^{nr(\mathbf{x})s(\mathbf{y}|\mathbf{x})}\notag\\
 & \quad= \frac{e^{-n(H_{r_{\boldsymbol{X}}s_{\boldsymbol{Y}|\boldsymbol{X}}}(\boldsymbol{Y}|\boldsymbol{X})+D(s_{\boldsymbol{Y}|\boldsymbol{X}}\| q_{\boldsymbol{Y}|\boldsymbol{X}}|r_{\boldsymbol{X}}))}}{|\boldsymbol{\mathcal{W}}|}\sum_{\mathbf{w}\in\boldsymbol{\mathcal{W}}}|\mathcal{A}(\mathbf{w};r,s)| \label{convProofEq7}
\end{align}
where (f) follows from the definition of $\mathcal{A}(\mathbf{w};r,s)$ in~\eqref{defSetA} and Definition~\ref{defDiscreteMemoryless}.
\subsection{Bounding the Size of $\mathcal{A}(\mathbf{w};r,s)$}
For each $r_{\boldsymbol{X}}\in\mathcal{P}_n(\boldsymbol{\mathcal{X}})$, each $s_{\boldsymbol{Y}|\boldsymbol{X}}\in \mathcal{P}_n(\boldsymbol{\mathcal{Y}}|r_{\boldsymbol{X}})$ and each $\mathbf{w}\in \boldsymbol{\mathcal{W}}$, since
\begin{align*}
\sum_{\mathbf{y}^n\in \mathcal{A}(\mathbf{w};r,s)} \prod_{k=1}^n s_{\boldsymbol{Y}|\boldsymbol{X}}(\mathbf{y}_k|\mathbf{x}_k(\mathbf{w},\mathbf{y}^{k-1})) \le 1,
\end{align*}
it follows that
\begin{align*}
\sum_{\mathbf{y}^n\in \mathcal{A}(\mathbf{w};r,s)} \prod_{\substack{\mathbf{x}, \mathbf{y}}} s_{\boldsymbol{Y}|\boldsymbol{X}}(\mathbf{y}|\mathbf{x})^{nr(\mathbf{x})s(\mathbf{y}|\mathbf{x})} \le 1
\end{align*}
(recall the definition of~$\mathcal{A}(\mathbf{w};r,s)$ in~\eqref{defSetA}),
which implies that
\begin{align*}
\sum_{\mathbf{y}^n\in \mathcal{A}(\mathbf{w};r,s)}  e^{-nH_{r_{\boldsymbol{X}}s_{\boldsymbol{Y}|\boldsymbol{X}}}(\boldsymbol{Y}|\boldsymbol{X})} \le 1,
\end{align*}
which then implies that
\begin{align}
|\mathcal{A}(\mathbf{w};r,s)| \le e^{nH_{r_{\boldsymbol{X}}s_{\boldsymbol{Y}|\boldsymbol{X}}}(\boldsymbol{Y}|\boldsymbol{X})}. \label{convProofEq8}
\end{align}
Combining~\eqref{convProofEq7} and~\eqref{convProofEq8}, we have for each $r_{\boldsymbol{X}}\in\mathcal{P}_n(\boldsymbol{\mathcal{X}})$ and each $s_{\boldsymbol{Y}|\boldsymbol{X}}\in \mathcal{P}_n(\boldsymbol{\mathcal{Y}}|r_{\boldsymbol{X}})$
\begin{align}
 &\frac{1}{|\boldsymbol{\mathcal{W}}|}\sum_{\mathbf{w}\in \boldsymbol{\mathcal{W}}}\sum\limits_{\substack{\mathbf{y}^n \in\\ \mathcal{A}(\mathbf{w}; r, s)}}\prod_{k=1}^n  p_{\boldsymbol{Y}_k|\boldsymbol{X}_k}(\mathbf{y}_k|\mathbf{x}_k(\mathbf{w}, \mathbf{y}^{k-1}))\times\mathbf{1}\left\{\hat{\mathbf{w}}=\mathbf{w}\right\} \le e^{-n D(s_{\boldsymbol{Y}|\boldsymbol{X}}\|q_{\boldsymbol{Y}|\boldsymbol{X}} |r_{\boldsymbol{X}})}. \label{convProofEq9}
 \end{align}
 \subsection{Relating the Bounds on Correct Decoding Probability to the Cut-Set Bound}
Defining
 \begin{equation}
 \alpha_T(r,s)\stackrel{\mathrm{def}}{=} e^{-n\big(\sum\limits_{(i,j)\in T\times T^c}R_{i,j}-I_{r_{\boldsymbol{X}}s_{Y_{T^c}|\boldsymbol{X}}}(X_T; Y_{T^c}|X_{T^c})+D(s_{Y_{T^c}|\boldsymbol{X}}\|q_{Y_{T^c}|\boldsymbol{X}} |r_{\boldsymbol{X}})\big)} \label{defAlphaT}
 \end{equation}
 and
 \begin{equation}
 \beta(r,s)\stackrel{\mathrm{def}}{=} e^{-n D(s_{\boldsymbol{Y}|\boldsymbol{X}}\|q_{\boldsymbol{Y}|\boldsymbol{X}} |r_{\boldsymbol{X}})}, \label{defBetaT}
 \end{equation}
 we obtain from~\eqref{convProofEq6} and~\eqref{convProofEq9} that for each $r_{\boldsymbol{X}}\in\mathcal{P}_n(\boldsymbol{\mathcal{X}})$ and each $s_{\boldsymbol{Y}|\boldsymbol{X}}\in \mathcal{P}_n(\boldsymbol{\mathcal{Y}}|r_{\boldsymbol{X}})$,
 \begin{align}
 &\frac{1}{|\boldsymbol{\mathcal{W}}|}\sum_{\mathbf{w}\in \boldsymbol{\mathcal{W}}}\sum\limits_{\substack{\mathbf{y}^n \in\\ \mathcal{A}(\mathbf{w}; r, s)}}\prod_{k=1}^n  p_{\boldsymbol{Y}_k|\boldsymbol{X}_k}(\mathbf{y}_k|\mathbf{x}_k(\mathbf{w}, \mathbf{y}^{k-1})) \times\mathbf{1}\left\{\hat{\mathbf{w}}=\mathbf{w}\right\} \le \min\{ \alpha_T(r,s), \beta(r,s)\}. \label{convProofEq10}
\end{align}
Combining \eqref{convProofEq1}, \eqref{convProofEq2} and~\eqref{convProofEq3} and using the fact that~\eqref{convProofEq10} holds for each $r_{\boldsymbol{X}}\in\mathcal{P}_n(\boldsymbol{\mathcal{X}})$, each $s_{\boldsymbol{Y}|\boldsymbol{X}}\in \mathcal{P}_n(\boldsymbol{\mathcal{Y}}|r_{\boldsymbol{X}})$ and any arbitrary non-empty $T\subsetneq\mathcal{I}$, we conclude that
\begin{align}
1-\varepsilon_n \le \sum_{r_{\boldsymbol{X}}\in\mathcal{P}_n(\boldsymbol{\mathcal{X}})}\sum_{s_{\boldsymbol{Y}|\boldsymbol{X}}\in \mathcal{P}_n(\boldsymbol{\mathcal{Y}}|r_{\boldsymbol{X}})} \min\{\alpha_{V_r}(r,s), \beta(r,s)\} \label{convProofEq11}
\end{align}
where the set $V_r\subseteq\mathcal{I}$ was carefully chosen to depend on $r_{\boldsymbol{X}}\in\mathcal{P}_n(\boldsymbol{\mathcal{X}})$ so that \eqref{convProofRate} holds. Note that~\eqref{convProofEq11} resembles \cite[Eq.~(7)]{CsiszarKorner82}.
Let $\xi>0$ be a positive constant to be specified later. It then follows from~\eqref{convProofEq11} that
\begin{align}
1-\varepsilon_n &\le \sum_{r_{\boldsymbol{X}}\in\mathcal{P}_n(\boldsymbol{\mathcal{X}})}\sum_{s_{\boldsymbol{Y}|\boldsymbol{X}}\in \mathcal{P}_n(\boldsymbol{\mathcal{Y}}|r_{\boldsymbol{X}})} \min\{\alpha_{V_r}(r,s), \beta(r,s)\} \notag\\
 &\qquad \times \left(\mathbf{1}\left\{D(s_{\boldsymbol{Y}|\boldsymbol{X}}\|q_{\boldsymbol{Y}|\boldsymbol{X}} |r_{\boldsymbol{X}})\ge \xi\right\}+\mathbf{1}\left\{D(s_{\boldsymbol{Y}|\boldsymbol{X}}\|q_{\boldsymbol{Y}|\boldsymbol{X}} |r_{\boldsymbol{X}})<\xi\right\}\right). \label{convProofEq12}
\end{align}
\subsection{Bounding the Correct Decoding Probability in Two Different Ways}
Recalling that $\delta>0$ was chosen such that \eqref{convProofRate} holds, we choose $\xi>0$ to be a positive constant such that the following statement holds  for all non-empty $T\subsetneq \mathcal{I}$:
\begin{equation}
|I_{g_{\boldsymbol{X}, \boldsymbol{Y}}}(X_T; Y_{T^c}|X_{T^c}) - I_{h_{\boldsymbol{X}, \boldsymbol{Y}}}(X_T; Y_{T^c}|X_{T^c})|\le \delta/2  \label{convProofEq13}
\end{equation}
for all distributions $g_{\boldsymbol{X}, \boldsymbol{Y}}$ and $h_{\boldsymbol{X},\boldsymbol{Y}}$ defined on $(\boldsymbol{\mathcal{X}}, \boldsymbol{\mathcal{Y}})$ that satisfy
\begin{equation*}
\|g_{\boldsymbol{X}, \boldsymbol{Y}}-h_{\boldsymbol{X}, \boldsymbol{Y}}\|_{\mathcal{L}_1} \le \sqrt{2\xi}.
\end{equation*}
The existence of such a $\xi>0$ is guaranteed by the fact that the mapping $p_{\boldsymbol{X}, \boldsymbol{Y}}\mapsto I_{p_{\boldsymbol{X}, \boldsymbol{Y}}}(X_T; Y_{T^c}|X_{T^c})$ is continuous with respect to the $\mathcal{L}_1$-distance for all non-empty $T\subsetneq \mathcal{I}$. Following~\eqref{convProofEq12}, we consider the following two chains of inequalities for each $r_{\boldsymbol{X}}\in\mathcal{P}_n(\boldsymbol{\mathcal{X}})$ and each $s_{\boldsymbol{Y}|\boldsymbol{X}}\in \mathcal{P}_n(\boldsymbol{\mathcal{Y}}|r_{\boldsymbol{X}})$:
\begin{align}
\min\{\alpha_{V_r}(r,s), \beta(r,s)\} \times \mathbf{1}\left\{D(s_{\boldsymbol{Y}|\boldsymbol{X}}\|q_{\boldsymbol{Y}|\boldsymbol{X}} |r_{\boldsymbol{X}})\ge \xi\right\}
& \le \beta(r,s)\times\mathbf{1}\left\{D(s_{\boldsymbol{Y}|\boldsymbol{X}}\|q_{\boldsymbol{Y}|\boldsymbol{X}} |r_{\boldsymbol{X}})\ge \xi\right\}\notag\\
& \stackrel{\eqref{defBetaT}}{\le} e^{-n\xi} \label{convProofEq14}
\end{align}
and
\begin{align}
&\min\{\alpha_{V_r}(r,s), \beta(r,s)\} \times \mathbf{1}\left\{D(s_{\boldsymbol{Y}|\boldsymbol{X}}\|q_{\boldsymbol{Y}|\boldsymbol{X}} |r_{\boldsymbol{X}})< \xi\right\} \notag\\
&\quad \stackrel{\text{(g)}}{\le} \alpha_{V_r}(r,s)\times\mathbf{1}\left\{\|r_{\boldsymbol{X}}s_{\boldsymbol{Y}|\boldsymbol{X}}-r_{\boldsymbol{X}}q_{\boldsymbol{Y}|\boldsymbol{X}} \|_{\mathcal{L}_1}< \sqrt{2\xi}\right\}\notag\\
& \quad\stackrel{\eqref{convProofEq13}}{\le} \alpha_{V_r}(r,s)\times\mathbf{1}\left\{|I_{r_{\boldsymbol{X}}s_{ \boldsymbol{Y}|\boldsymbol{X}}}(X_{V_r}; Y_{{V_r}^c}|X_{{V_r}^c}) - I_{r_{\boldsymbol{X}}q_{ \boldsymbol{Y}|\boldsymbol{X}}}(X_{V_r}; Y_{{V_r}^c}|X_{{V_r}^c})|\le \delta/2\right\}\notag\\
&\quad \stackrel{\eqref{defAlphaT}}{\le}e^{-n\big(\sum\limits_{(i,j)\in {V_r}\times {V_r}^c}R_{i,j}-I_{r_{\boldsymbol{X}}s_{Y_{{V_r}^c}|\boldsymbol{X}}}(X_{V_r}; Y_{{V_r}^c}|X_{{V_r}^c})\big)}\notag\\
&\quad\qquad \times\mathbf{1}\left\{|I_{r_{\boldsymbol{X}}s_{ \boldsymbol{Y}|\boldsymbol{X}}}(X_{V_r}; Y_{{V_r}^c}|X_{{V_r}^c}) - I_{r_{\boldsymbol{X}}q_{ \boldsymbol{Y}|\boldsymbol{X}}}(X_{V_r}; Y_{{V_r}^c}|X_{{V_r}^c})|\le \delta/2\right\}\notag \\
&\quad \le e^{-n\big(\sum\limits_{(i,j)\in {V_r}\times {V_r}^c}R_{i,j}-I_{r_{\boldsymbol{X}}q_{Y_{{V_r}^c}|\boldsymbol{X}}}(X_{V_r}; Y_{{V_r}^c}|X_{{V_r}^c})-\delta/2\big)} \notag\\
&\quad \stackrel{\eqref{convProofRate}}{\le} e^{-n\delta/2}, \label{convProofEq15}
\end{align}
where (g) follows from Pinsker's inequality.
Combining \eqref{convProofEq12}, \eqref{convProofEq14} and \eqref{convProofEq15} followed by using the fact due to~\eqref{typeClassBound} that
\begin{equation*}
|\mathcal{P}_n(\boldsymbol{\mathcal{X}} \times \boldsymbol{\mathcal{Y}})| \le (n+1)^{|\boldsymbol{\mathcal{X}}||\boldsymbol{\mathcal{Y}}|},
\end{equation*}
 we obtain
\begin{align}
1-\varepsilon_n &\le (n+1)^{|\boldsymbol{\mathcal{X}}||\boldsymbol{\mathcal{Y}}|}e^{-n\min\{\xi,\delta/2\}} \label{convProofEq16}
\end{align}
(analogous to the last inequality in~\cite{CsiszarKorner82}),
which implies~\eqref{convProofError} as $|\boldsymbol{\mathcal{X}}||\boldsymbol{\mathcal{Y}}|$, $\xi$ and $\delta$ are positive constants that do not depend on~$n$.
Since~\eqref{convProofError} holds for any sequence of $(n, \mathbf{R}, \varepsilon_n)$-codes with $\mathbf{R} \notin \mathcal{R}_{\mathrm{cut-set}}$, it follows that~\eqref{convProofSt} holds for all $\varepsilon\in[0,1)$.

\section{Preliminaries for Proving Theorem~\ref{AWGN:thmMainResult} --- Gaussian Types} \label{AWGN:sectionPrelim}
In this section, we generalize the definitions and results of the method of types \cite[Ch.~2]{Csi97} to the Gaussian case. Our generalization is inspired by the previous generalizations to the Gaussian case with two variables for the guessing problem in~\cite[Sec.~VI]{ArikanMerhav98} and with three variables for the source coding problem in~\cite[Appendix D]{KellyWagner11}. More specifically, we generalize the method of types to the multivariate case in order to investigate the channel coding problem for any Gaussian network. Throughout this section, we let $n$ denote an arbitrary natural number, and let~$T$, $T_1$ and $T_2$ denote any arbitrary non-empty subsets of $\mathcal{I}$.
\subsection{Gaussian Types} \label{AWGN:subsecGaussianType}
\begin{Definition}\label{AWGN:defCorrelationMatrix}
The \textit{empirical correlation} between two sequences of column vectors $x_T^n \in \mathbb{R}^{n|T| }$ and $y_T^n \in \mathbb{R}^{n|T| }$ is the~$|T|\times |T|$ matrix defined as
\begin{equation}
\Upsilon^{[x_T^n, y_T^n]} \stackrel{\mathrm{def}}{=} \frac{1}{n} \sum_{k=1}^nx_{T,k}\,y_{T,k}^t. \label{AWGN:defGaussianType}
\end{equation}
The \textit{autocorrelation} of a column vector $x_T \in \mathbb{R}^{|T|}$ is defined as
\begin{equation}
\R{[x_T]} \stackrel{\mathrm{def}}{=} \Upsilon^{[x_T, x_T]} = x_T x_T^t. \label{AWGN:defAutoCorr}
\end{equation}
The \textit{empirical autocorrelation} of a sequence of column vectors $x_T^n \in \mathbb{R}^{n|T|}$ is defined as
\begin{equation*}
\R{[x_T^n]} \stackrel{\mathrm{def}}{=} \Upsilon^{[x_T^n, x_T^n]} = \frac{1}{n}\sum_{k=1}^n \R{[x_{T,k}]}.
\end{equation*}
\end{Definition}

\smallskip
\begin{Definition}\label{AWGN:defType}
The \textit{Gaussian type} of $(x_{T_1}^n, y_{T_2}^n)\in \mathbb{R}^{n|T_1|}\times \mathbb{R}^{n|T_2|}$ is the $(|T_1|+ |T_2|)\times (|T_1|+ |T_2|)$ matrix
\begin{equation}
K^{[x_{T_1}^n,y_{T_2}^n]}\stackrel{\mathrm{def}}{=}\left[\begin{matrix} \Upsilon^{[x_{T_1}^n, x_{T_1}^n]} & \Upsilon^{[x_{T_1}^n, y_{T_2}^n]} \\ \Upsilon^{[y_{T_2}^n, x_{T_1}^n]} & \Upsilon^{[y_{T_2}^n, y_{T_2}^n]}\end{matrix}\right]. \label{AWGN:defKCov}
\end{equation}
\end{Definition}
\smallskip

For any given $(n, \mathbf{R}, \mathbf{P})$-code which induces the probability distribution $p_{\boldsymbol{X}}^n$, the almost-sure power constraints~\eqref{AWGN:peakPowerConstraints} imply that
 \begin{equation}
\int_{\mathbb{R}^{n}}p_{X_i^n}(x_i^n)\times\mathbf{1}\left\{\frac{1}{n}\sum_{k=1}^n x_{i,k}^2\le P_i\right\}\mathrm{d}x_i^n =1 \label{AWGN:typeXnProb1*}
 \end{equation}
 for all $i\in\mathcal{I}$, which implies by the definition of $\mathcal{S}(\mathbf{P})$ in~\eqref{AWGN:defSetS} that the probability that the empirical autocorrelation of~$\boldsymbol{X}^n$ falling inside~$\mathcal{S}(\mathbf{P})$ is~$1$, i.e.,
\begin{align}
\int_{\mathbb{R}^{nN}}p_{\boldsymbol{X}^n}(\mathbf{x}^n)\times\mathbf{1}\left\{\R{[\mathbf{x}^n]}\in \mathcal{S}(\mathbf{P})\right\}\mathrm{d}\mathbf{x}^n =1. \label{AWGN:typeXnProb1}
\end{align}
For each $\delta>0$ and each $N\times N$ matrix $A\in \mathbb{R}^{N\times N}$, define the $\delta$-neighborhood of~$A$ as
  \begin{equation}
  \Gamma_\delta(A)\stackrel{\mathrm{def}}{=} \left\{B\in\mathbb{R}^{N\times N}\left|\,\, -\, \delta\cdot 1^{N\times N} \le B - A \le \delta\cdot 1^{N\times N}\right.\right\}. \label{defSetGamma}
  \end{equation}
Let
\begin{equation}
\mathcal{U}_{\boldsymbol{X}, \boldsymbol{Y}}^{(\delta, \mathbf{P})} \stackrel{\mathrm{def}}{=} \left\{\left[
\begin{matrix}
K^{11}&K^{12}\vspace{0.04 in}\\K^{21} &K^{22} \end{matrix}
\right]\in \mathbb{R}^{2N\times 2N}\left|\,
\parbox[c]{2.65 in}{$ K^{11} \in \mathcal{S}(\mathbf{P}) $,\vspace{0.04 in}\\
$ K^{12}- K^{11}\mathbf{G}^t\in \Gamma_\delta(0^{N\times N})$,\vspace{0.04 in}\\
 $  K^{21}-\mathbf{G}K^{11}\in \Gamma_\delta(0^{N\times N})$,\vspace{0.04 in}\\$K^{22} + \mathbf{G}K^{11}\mathbf{G}^t - \mathbf{G}K^{12}- K^{21}\mathbf{G}^t\in \Gamma_\delta(\mathbf{\Sigma})$}
\right.\right\} \label{AWGN:defSetU}
\end{equation}
be a collection of typical Gaussian types of $(\mathbf{x}^n, \mathbf{y}^n)$ where the empirical autocorrelation of $\mathbf{x}^n$ falls inside $\mathcal{S}(\mathbf{P})$ and the empirical autocorrelation of $\mathbf{z}^n$ falls into some neighborhood of the noise covariance matrix~$\mathbf{\Sigma}$.
The following lemma shows that the probability that the Gaussian type of $(\boldsymbol{X}^n, \boldsymbol{Y}^n)$ falls outside $\mathcal{U}_{\boldsymbol{X}, \boldsymbol{Y}}^{(\delta, \mathbf{P})}$ is exponentially small.  The proof of Lemma~\ref{AWGN:lemmaJointTypeExponentiallySmall} is tedious, hence is deferred to Appendix~\ref{AWGN:appendixB}.
\smallskip
\begin{Lemma} \label{AWGN:lemmaJointTypeExponentiallySmall}
For any $\delta>0$, there exists a constant $\tau>0$ which is a function of $(\mathbf{P},\mathbf{\Sigma})$ such that for all sufficiently large~$n$,
\begin{align*}
\int_{\mathbb{R}^{nN}}\int_{\mathbb{R}^{nN}}p_{\boldsymbol{X}^n, \boldsymbol{Y}^n}(\mathbf{x}^n,\mathbf{y}^n)\times\mathbf{1}\left\{K^{[\mathbf{x}^n,\mathbf{y}^n]}\in \mathcal{U}_{\boldsymbol{X}, \boldsymbol{Y}}^{(\delta, \mathbf{P})}\right\}\mathrm{d}\mathbf{y}^n\mathrm{d}\mathbf{x}^n > 1-e^{-\tau n}
\end{align*}
holds for any $(n, \mathbf{R}, \mathbf{P})$-code where $p_{\boldsymbol{X}^n, \boldsymbol{Y}^n}$ is the distribution induced by the code.
\end{Lemma}
\subsection{Quantizers, Types and Type Classes} \label{AWGN:subsecQuantize}
In Definition~\ref{AWGN:defType}, we have defined the Gaussian type of a given sequence. However, there are uncountably many Gaussian types. Therefore, we would like to quantize Euclidean space uniformly so that the quantization error along each dimension is less than $\Delta$. To this end, we define $\Delta$-quantizers in Definition~\ref{AWGN:definitionTypeRep}, which will be used to approximate any covariance matrix within $\Delta$ units along each dimension.
\smallskip
\begin{Definition}\label{AWGN:definitionTypeRep}
Fix any positive number $\Delta$. An $N\times N$ real-valued matrix $\Lambda$ is called a \emph{$\Delta$-quantizer} if there exists an $N\times N$ matrix $\Pi$ whose elements are integers such that
\begin{align*}
\Lambda = \Delta\Pi.
\end{align*}
The set of $\Delta$-quantizers is denoted by $\mathcal{L}^\Delta$, which can be viewed as a scaled version of the $N^2$-dimensional integer lattice.
\end{Definition}

\begin{Definition}\label{AWGN:definitionDeltaQuantizer}
Given any $\Delta$-quantizer $\Lambda\in \mathcal{L}^\Delta$, the \emph{$\Delta$-box represented by $\Lambda$} is defined as
\begin{align*}
\mathcal{V}_{\Lambda}^{\Delta} \stackrel{\mathrm{def}}{=} \left\{\left. B\in \mathbb{R}^{N\times N} \right|\Lambda\le B<\Lambda+ \Delta  \cdot 1^{N\times N} \right\}.
\end{align*}
A set $\mathcal{V}$ is called a \emph{$\Delta$-box} if it is a $\Delta$-box represented by some $\Lambda$ in $\mathcal{L}^\Delta$.
\end{Definition}
\smallskip
By Definition~\ref{AWGN:definitionDeltaQuantizer}, we can see that the size of $\mathcal{L}^\Delta$ is countably infinite and the set $\{\mathcal{V}_{\Lambda}^{\Delta}:\Lambda\in \mathcal{L}^\Delta\}$ forms a partition on $\mathbb{R}^{N\times N}$. Recalling the definition of $\mathcal{S}(\mathbf{P})$ in~\eqref{AWGN:defSetS}, we define for each $\gamma> 0$ the set of positive definite covariance matrices
\begin{equation}
\mathcal{S}_{\gamma}(\mathbf{P}) \stackrel{\mathrm{def}}{=} \left\{ \mathbf{K}\in \mathcal{S}(\mathbf{P}) \left|\, \text{For all non-empty $T\subseteq \mathcal{I}$, all the eigenvalues of $K_{T\times T}$ are at least $\gamma$}\right.\right\}. \label{AWGN:defSetSdelta}
\end{equation}
In the following definition, we define a subset of $\mathcal{L}^\Delta$ called the set of input $(\Delta, \gamma, \mathbf{P})$-quantizers, denoted by $\mathcal{L}^{(\Delta, \gamma, \mathbf{P})}$, so that the size of $\mathcal{L}^{(\Delta, \gamma, \mathbf{P})}$ is finite.
\smallskip
\begin{Definition}\label{AWGN:definitionType}
The \emph{set of input $(\Delta, \gamma, \mathbf{P})$-quantizers} is defined as
\begin{align}
\mathcal{L}^{(\Delta, \gamma, \mathbf{P})}\stackrel{\mathrm{def}}{=} \left\{ \Lambda \in \mathcal{L}^\Delta \left| \mathcal{V}_{\Lambda}^{\Delta}\cap \mathcal{S}_\gamma(\mathbf{P})\ne \emptyset\right.\right\}. \label{AWGN:defDeltaInputType}
\end{align}
\end{Definition}
Definition~\ref{AWGN:definitionType} implies that
\begin{equation}
\bigcup\limits_{\Lambda \in \mathcal{L}^{(\Delta, \gamma, \mathbf{P})}} \mathcal{V}_{\Lambda}^{\Delta} \supseteq \mathcal{S}_\gamma(\mathbf{P}). \label{AWGN:eqnTypeCovering}
\end{equation}
The following proposition shows that $|\mathcal{L}^{(\Delta, \gamma, \mathbf{P})}|$ is finite, whose proof is simple and is given in Appendix~\ref{AWGN:appendixC} for the sake of completeness.
\smallskip
\begin{Proposition} \label{AWGN:propositionCardinalityBound}
For any $\Delta>0$ and any $\gamma>0$, we have
\begin{equation*}
|\mathcal{L}^{(\Delta, \gamma, \mathbf{P})}| \le \prod\limits_{(i,j)\in \mathcal{I}\times \mathcal{I}}\left(2\left\lceil\frac{\sqrt{P_i P_j}}{\Delta}\right\rceil+1\right).
\end{equation*}
\end{Proposition}

We are ready to construct $(\Delta, \gamma, \mathbf{P})$-types as follows.
\smallskip
\begin{Definition}\label{AWGN:definitionDeltaType}
For each $\Lambda\in \mathcal{L}^{(\Delta, \gamma, \mathbf{P})}$, choose and fix one covariance matrix $K_{\Lambda}\in \mathcal{V}_{\Lambda}^{\Delta}\cap \mathcal{S}_\gamma(\mathbf{P})$ and call it the \emph{$(\Delta, \gamma, \mathbf{P})$-type represented by $\Lambda$}.
 A covariance matrix $J\in \mathcal{S}(\mathbf{P})$ is called a \emph{$(\Delta, \gamma, \mathbf{P})$-type} if it is a $(\Delta, \gamma, \mathbf{P})$-type represented by $\Lambda$ for some $\Lambda\in \mathcal{L}^{(\Delta, \gamma, \mathbf{P})}$, and we let
 \begin{equation*}
 \mathcal{V}^\Delta(J) \stackrel{\mathrm{def}}{=} \mathcal{V}_{\Lambda}^{\Delta}
 \end{equation*}
 be the $\Delta$-box that contains $J$. The set of $(\Delta, \gamma, \mathbf{P})$-type is denoted by $\mathcal{P}^{(\Delta, \gamma, \mathbf{P})}$.
\end{Definition}
\smallskip

The following corollary follows directly from Definition~\ref{AWGN:definitionDeltaType}, \eqref{AWGN:eqnTypeCovering} and Proposition~\ref{AWGN:propositionCardinalityBound}, hence the proof is omitted.
\smallskip
\begin{Corollary} \label{AWGN:corollaryCardinalityBound}
For any $\Delta>0$ and $\gamma>0$,
\begin{equation}
\bigcup\limits_{J\in \mathcal{P}^{(\Delta, \gamma, \mathbf{P})}} \mathcal{V}^{\Delta}(J) \supseteq \mathcal{S}_\gamma(\mathbf{P}). \label{AWGN:eqnTypeCovering*}
\end{equation}
In addition,
\begin{equation*}
|\mathcal{P}^{(\Delta, \gamma, \mathbf{P})}| \le \prod\limits_{(i,j)\in \mathcal{I}\times \mathcal{I}}\left(2\left\lceil\frac{\sqrt{P_i P_j}}{\Delta}\right\rceil+1\right).
\end{equation*}
\end{Corollary}
\begin{Definition} \label{AWGN:definitionJointTypeClass}
Fix any $n\in\mathbb{N}$. For any $(\Delta, \gamma, \mathbf{P})$-type $J \in \mathcal{P}^{(\Delta, \gamma, \mathbf{P})}$, the \textit{input $\Delta$-type class of $J$} is defined as
\begin{equation*}
\mathcal{T}_{J}^{(n, \Delta)}(\boldsymbol{X}) \stackrel{\mathrm{def}}{=} \left\{\mathbf{x}^n\in \mathbb{R}^{nN} \left|\, \R{[\mathbf{x}^n]}  \in  \mathcal{V}^{\Delta}(J) \right.\right\}.
\end{equation*}
In addition, the \textit{joint $(\Delta, \delta, \mathbf{P})$-type class of $J$} is defined as
\begin{equation}
\mathcal{T}_{J}^{(n, \Delta, \delta, \mathbf{P})}(\boldsymbol{X}, \boldsymbol{Y}) \stackrel{\mathrm{def}}{=} \left\{(\mathbf{x}^n, \mathbf{y}^n)\in \mathbb{R}^{nN} \times  \mathbb{R}^{nN}  \left|\, \mathbf{x}^n  \in  \mathcal{T}_{J}^{(n, \Delta)}(\boldsymbol{X}), K^{[\mathbf{x}^n, \mathbf{y}^n]}  \in  \mathcal{U}_{\boldsymbol{X}, \boldsymbol{Y}}^{(\delta, \mathbf{P})} \right.\right\},  \label{AWGN:defJointTypeClass}
\end{equation}
and the \textit{joint $(\Delta, \delta, \mathbf{P})$-type class of $J$ restricted to $(X_{T^c}, Y_{T^c})$} is defined as
\begin{align}
\mathcal{T}_{J}^{(n, \Delta, \delta, \mathbf{P})}(X_{T^c}, Y_{T^c}) \stackrel{\mathrm{def}}{=}\left\{(x_{T^c}^n, y_{T^c}^n)\in\mathbb{R}^{n|T^c|} \times  \mathbb{R}^{n|T^c|}  \left|\, \parbox[c]{2.9 in}{There exists a pair $(\bar{x}_\mathcal{I}^n, \bar{y}_\mathcal{I}^n)\in \mathcal{T}_{J}^{(n, \Delta, \delta, \mathbf{P})}(\boldsymbol{X},\boldsymbol{Y})$ such that $(\bar x_{T^c}^n, \bar y_{T^c}^n) = (x_{T^c}^n,y_{T^c}^n)$} \right.\right\}. \label{AWGN:defJointTypeClassRestricted}
\end{align}
\end{Definition}
\smallskip

The proof of Theorem~\ref{AWGN:thmMainResult} involves the following two simple but useful bounds.
\smallskip
\begin{Proposition} \label{AWGN:propositionInverseBound}
Suppose $K\succ 0$ is an $N\times N$ real-valued matrix. Let $k_{\min}>0$ be the smallest eigenvalue of~$K$. Then, we have
\begin{align*}
K^{-1}\in \Gamma_{\frac{N}{k_{\min}}}(0^{N\times N})
\end{align*}
where the definition of $\Gamma_{\delta}(0^{N\times N})$ is given in~\eqref{defSetGamma}.
\end{Proposition}
\begin{proof}
The desired result can be obtained by diagonalizing $K$. More precisely, let
\begin{equation}
K = U D U^t \label{eqnDiagonalize}
\end{equation}
be the eigendecomposition of $K$, where $U$ is a unitary matrix whose rows comprise an orthonormal basis of eigenvectors of $K$ and $D$ is a diagonal matrix with positive diagonal elements $\lambda_1, \lambda_2, \ldots, \lambda_N$ satisfying
$
\lambda_1\le \lambda_2 \le \ldots \le \lambda_N$.
 Let $k_{\min}\stackrel{\mathrm{def}}{=} \lambda_1>0$. Inverting both sides of~\eqref{eqnDiagonalize} followed by straightforward multiplications  reveals that $K^{-1}=U D^{-1} U^t$. Since the largest value of the diagonal matrix $D^{-1}$ equals $1/k_{\min}$ and the magnitudes of the elements in $U$ are no larger than~$1$ (the rows of~$U$ are orthonormal), it follows by inspecting $K^{-1}=U D^{-1} U^t$ that the magnitudes of the elements in $K^{-1}$ are no larger than~$\frac{N}{k_{\min}}$.
\end{proof}
\smallskip
\begin{Proposition} \label{AWGN:propositionProductBound}
Suppose $\Pi_1$ and $\Pi_2$ are $N_1\times N_2$ and $N_2\times N_3$ real-valued matrices respectively. Let
\begin{equation*}
\pi^{\max}_i=\max\{|r|\, : \text{$r$ is an entry in~$\Pi_i$}\}
\end{equation*}
 for each $i\in\{1,2\}$. Then, we have
\begin{align*}
\Pi_1\Pi_2\in \Gamma_{N_2 \pi^{\max}_1 \pi^{\max}_2}(0^{N_1\times N_3}).
\end{align*}
\end{Proposition}
\begin{proof}
The desired result can be obtained by using the facts that $\Pi_1 \in \Gamma_{\pi^{\max}_1}(0^{N_1\times N_2})$ and $\Pi_2 \in \Gamma_{\pi^{\max}_2}(0^{N_2\times N_3})$.
\end{proof}
\smallskip

In the proof of Theorem~\ref{AWGN:thmMainResult}, a key step involves the following lemma which bounds the product probabilities $\prod_{k=1}^n q_{Y_{T^c}|\boldsymbol{X}}(y_{T^c,k}|\mathbf{x}_k)$ for each $(\Delta, \gamma, \mathbf{P})$-type $J \in \mathcal{P}^{(\Delta, \gamma, \mathbf{P})}$ and each $(\mathbf{x}^n, \mathbf{y}^n)\in \mathcal{T}_{J}^{(n, \Delta, \delta, \mathbf{P})}(\boldsymbol{X}, \boldsymbol{Y})$.
 Since the proof of Lemma~\ref{AWGN:lemmaProductProbBound} is tedious, it is relegated to Appendix~\ref{AWGN:appendixD}.
 \smallskip
\begin{Lemma} \label{AWGN:lemmaProductProbBound}
Let $\sigma_{\min}$ be the smallest eigenvalue of~$\mathbf{\Sigma}$. Fix any $T\subsetneq \mathcal{I}$, and fix any $(\Delta, \gamma, \mathbf{P})$-type $J \in \mathcal{P}^{(\Delta, \gamma, \mathbf{P})}$. Then for each $(\mathbf{x}^n, \mathbf{y}^n)\in \mathcal{T}_{J}^{(n, \Delta, \delta, \mathbf{P})}(\boldsymbol{X}, \boldsymbol{Y})$, we have
\begin{align}
\prod_{k=1}^n q_{Y_{T^c}|\boldsymbol{X}}(y_{T^c,k}|\mathbf{x}_k) \le e^{-n\left(\frac{1}{2}\log \left((2\pi e)^{|T^c|} |\Sigma_{T^c\times T^c}|\right)- \frac{\delta N^3}{2\sigma_{\min}} \right)}.  \label{AWGN:stLemmaProductProbBound}
\end{align}
\end{Lemma}

\section{Proof of Theorem~\ref{AWGN:thmMainResult}}\label{AWGN:secConverse}
In this section, we will show that
\begin{equation}
\mathcal{C}_\varepsilon \subseteq  \mathcal{R}_{\mathrm{cut-set}} \label{AWGN:convProofSt}
\end{equation}
 for all $\varepsilon\in[0,1)$ where $\mathcal{R}_{\mathrm{cut-set}}$ is as defined in~\eqref{AWGN:Rcutset}. It suffices to show that for any $\mathbf{R} \notin \mathcal{R}_{\mathrm{cut-set}}$ and any sequence of $(\bar n, \mathbf{R}, \mathbf{P}, \varepsilon_{\bar n})$-codes, the limit of the error probabilities must satisfy
 \begin{equation}
\lim_{\bar n\rightarrow \infty}\varepsilon_{\bar n} =1. \label{AWGN:convProofError}
 \end{equation}
 To this end, we fix a rate vector~$\mathbf{R} \notin \mathcal{R}_{\mathrm{cut-set}}$ and a sequence of $(\bar n, \mathbf{R}, \mathbf{P}, \varepsilon_{\bar n})$-codes.
  \subsection{Relating $\mathbf{R}$ to the Cut-Set Bound}
  Since $\mathbf{R} \notin \mathcal{R}_{\mathrm{cut-set}}$ and $\mathcal{R}_{\mathrm{cut-set}}$ is closed, it follows by the definition of~$\mathcal{R}_{\mathrm{cut-set}}$ in~\eqref{AWGN:Rcutset} that we can always find a positive number $\eta>0$ such that for any covariance matrix $\mathbf{K}\in \mathcal{S}(\mathbf{P})$, there exists a non-empty $V_{\mathbf{K}}\subsetneq\mathcal{I}$ that satisfies
\begin{equation*}
\sum\limits_{(i,j)\in V\times V^c} R_{i,j}
 \ge  \frac{1}{2}\log\left|I_{|V^c|} + G_{V^c\times V}K_{V|V^c} G_{V^c\times V}^t \left(\Sigma_{V^c\times V^c}\right)^{-1}\right|+\eta
 \end{equation*}
 where the shorthand notation $V$ is used to denote $V_{\mathbf{K}}$ and $K_{V|V^c}$ is as defined in~\eqref{AWGN:conditionalVar}.
Define
 \begin{equation}
 \eta(\delta)\stackrel{\mathrm{def}}{=} \frac{\delta N^2}{2\sigma_{\min}}\left(\delta N + (2Ng_{\max}+1)\delta + \frac{2N^4 g_{\max} (1+\delta)P_{\max}}{P_{\min}}+1\right) \label{AWGN:defEta}
 \end{equation}
  where $P_{\min}\stackrel{\mathrm{def}}{=} \min_{i\in\mathcal{I}}P_i >0$, $P_{\max}\stackrel{\mathrm{def}}{=} \max_{i\in\mathcal{I}}P_i >0$ and
  $\sigma_{\min}>0$ is defined as the smallest eigenvalue of~$\mathbf{\Sigma}$.
Then, we can always find a sufficiently small number $\delta >0$ such that for any covariance matrix $\mathbf{K}\in \mathcal{S}((1+\delta)\mathbf{P})$, the following inequality holds:
\begin{equation}
\sum\limits_{ (i,j)\in V\times V^c} (1-\delta)R_{i,j}
 \ge  \frac{1}{2}\log\left|I_{|V^c|} + G_{V^c\times V} K_{V|V^c}G_{V^c\times V}^t \left(\Sigma_{V^c\times V^c}\right)^{-1}\right|+2\eta(\delta). \label{AWGN:convProofRate*}
 \end{equation}
In particular, for any $\mathbf{K}\succ 0$, $K_{V|V^c}$ in~\eqref{AWGN:convProofRate*} admits the closed-form expression
\begin{equation}
K_{V|V^c}=K_{V\times V}-K_{V\times V^c}(K_{V^c\times V^c})^{-1}K_{V^c\times V} \label{AWGN:closedFormK}
\end{equation}
by the conditional variance formula for multivariate normal distributions in~\cite[Sec.~8.1.3]{IMM2012-03274}.

\subsection{Appending $N$ Redundant Transmissions}
In this proof, the quantity $K_{V|V^c}$ in~\eqref{AWGN:convProofRate*} is closely related to~$\R[\boldsymbol{X}^n]$, i.e., the empirical autocorrelation of $\boldsymbol{X}^n$. Since $K_{V|V^c}$ has a simple closed-form expression~\eqref{AWGN:closedFormK} if $\mathbf{K}\succ 0$, we are going to carefully append~$N$ redundant transmissions to every $(\bar n, \mathbf{R}, \mathbf{P}, \varepsilon_{\bar n})$-code so that $\R[\boldsymbol{X}^n]\succ 0$ holds with probability~1 for the resultant length-$(\bar n + N)$ code. To this end, we consider
each sufficiently large~$\bar n$ that satisfies
\begin{equation}
\bar nR_{i,j} \ge (\bar n+N)(1-\delta)R_{i,j} \label{AWGN:slightlySmallerMsgSize}
 \end{equation}
 for all $(i,j)\in\mathcal{I}\times \mathcal{I}$ and the corresponding $(\bar n, \mathbf{R}, \mathbf{P}, \varepsilon_{\bar n})$-code which has been fixed above, and construct an $(\bar n + N, (1-\delta)\mathbf{R}, (1+\delta)\mathbf{P}, \varepsilon_{\bar n})$-code as follows. In the first $\bar n$ time slots, the $(\bar n + N, (1-\delta)\mathbf{R} , (1+\delta)\mathbf{P}, \varepsilon_{\bar n})$-code is identical to the $(\bar n, \mathbf{R}, \mathbf{P}, \varepsilon_{\bar n})$-code. In the last $N$ time slots, the $N$ nodes transmit redundant information sequentially in this manner: In the $i^{\text{th}}$ last time slot for each $i\in\{1, 2, \ldots, N\}$, only node~$i$ transmits the non-zero symbol $\sqrt{\delta (\bar n + N) P_{\min}}$.
Since the empirical autocorrelation of every transmitted $\mathbf{x}^{\bar n}$ has a minimum eigenvalue of zero, the $N$ redundant information ensures that the empirical autocorrelation of every transmitted $\mathbf{x}^{\bar n +N}$ has a minimum eigenvalue of $\delta P_{\min}$.

 To simplify notation, let $n\stackrel{\mathrm{def}}{=}\bar n + N$, $\varepsilon_n \stackrel{\mathrm{def}}{=} \varepsilon_{\bar n}$ and $\mathbf{P}^{(\delta)}\stackrel{\mathrm{def}}{=} (1+\delta)\mathbf{P}$. For each $(n, (1-\delta)\mathbf{R}, \mathbf{P}^{(\delta)}, \varepsilon_{n})$-code constructed above, let $p_{\boldsymbol{W} , \boldsymbol{X}^n, \boldsymbol{Y}^n, \hat{\boldsymbol{W}} }$ be the induced probability distribution. By~\eqref{AWGN:defSetSdelta} and the construction above, we have for each $i\in\mathcal{I}$
 \begin{equation*}
\int_{\mathbb{R}^{nN}}p_{\boldsymbol{X}^n}(\mathbf{x}^n)\times\mathbf{1}\left\{\R{[\mathbf{x}^n]}\in \mathcal{S}_{\delta P_{\min}}(\mathbf{P}^{(\delta)})\right\}\mathrm{d}\mathbf{x}^n =1,
 \end{equation*}
which implies that $\R[\boldsymbol{X}^n]\succ 0$ holds with probability~1 for the $(n, (1-\delta)\mathbf{R}, \mathbf{P}^{(\delta)}, \varepsilon_{n})$-code.

\subsection{Simplifying the Correct Decoding Probability by Using the Memoryless Property}
Fix a sufficiently large~$n$ such that~\eqref{AWGN:slightlySmallerMsgSize} holds,
 \begin{align*}
\int_{\mathbb{R}^{nN}}\int_{\mathbb{R}^{nN}}p_{\boldsymbol{X}^n, \boldsymbol{Y}^n}(\mathbf{x}^n,\mathbf{y}^n)\times\mathbf{1}\left\{K^{[\mathbf{x}^n,\mathbf{y}^n]}\in \mathcal{U}_{\boldsymbol{X}, \boldsymbol{Y}}^{(\delta^2, \mathbf{P}^{(\delta)})}\right\}\mathrm{d}\mathbf{y}^n\mathrm{d}\mathbf{x}^n > 1-e^{-\tau n} 
\end{align*}
holds for some $\tau>0$ as a consequence of Lemma~\ref{AWGN:lemmaJointTypeExponentiallySmall} (where $\delta^2$ is chosen deliberately), and
\begin{align}
\frac{1}{n}\left( N^2 g_{\max}^2  + 2N^5 \left(\frac{g_{\max} (1+\delta)P_{\max}}{\delta P_{\min} }\right) + N^8 \left(\frac{g_{\max} (1+\delta)P_{\max}}{\delta P_{\min} } \right)^2\right) \le \delta \label{AWGN:sufficientlyLargeNInProof}
\end{align}
holds where
\begin{equation*}
g_{\max}\stackrel{\mathrm{def}}{=}\max\{|g| : \text{$g$ is an entry in $\mathbf{G}$}\}.
 \end{equation*}
 Unless specified otherwise, the probabilities are evaluated according to~$p_{\boldsymbol{W} , \boldsymbol{X}^n, \boldsymbol{Y}^n, \hat{\boldsymbol{W}} }$ in the rest of the proof.
By Lemma~\ref{AWGN:lemmaJointTypeExponentiallySmall} and the union bound, the probability of correct decoding can be bounded above as
\begin{align}
&1-\varepsilon_n \notag\\
&\quad= \Pr\left\{\bigcap_{i\in \mathcal{I}} \left\{\varphi_i\left(W_{\{i\}\times\mathcal{I}},Y_i^n\right) = W_{\mathcal{I}\times\{i\}}\right\}\right\} \notag\\
& \quad\le \Pr\left\{\bigcap_{i\in \mathcal{I}} \left\{\varphi_i\left(W_{\{i\}\times\mathcal{I}},Y_i^n\right) = W_{\mathcal{I}\times\{i\}}\right\}\cap\left\{K^{[\boldsymbol{X}^n, \boldsymbol{Y}^n]}\in \mathcal{U}_{\boldsymbol{X}, \boldsymbol{Y}}^{(\delta^2, \mathbf{P}^{(\delta)})}\right\}\cap\left\{\R[\boldsymbol{X}^n] \in \mathcal{S}_{\delta P_{\min}}(\mathbf{P}^{(\delta)})\right\}\right\} + e^{-\tau n}\notag \\
& \quad= \frac{1}{|\boldsymbol{\mathcal{W}}|}\sum_{\mathbf{w}\in \boldsymbol{\mathcal{W}}}\Pr\left\{\left. \parbox[c]{3.15 in}{$ \bigcap\limits_{i\in \mathcal{I}} \left\{\varphi_i\left(w_{\{i\}\times\mathcal{I}},Y_i^n\right) = w_{\mathcal{I}\times\{i\}}\right\}\\\cap\left\{K^{[\boldsymbol{X}^n, \boldsymbol{Y}^n]}\in \mathcal{U}_{\boldsymbol{X}, \boldsymbol{Y}}^{(\delta^2, \mathbf{P}^{(\delta)})}\right\}\cap\left\{\R[\boldsymbol{X}^n] \in \mathcal{S}_{\delta P_{\min}}(\mathbf{P}^{(\delta)})\right\}$}\right|\boldsymbol{W}=\mathbf{w}\right\} + e^{-\tau n}. \label{AWGN:convProofEq1}
\end{align}
In order to simplify notation, we define $
\hat w_{\mathcal{I}\times \{i\}}$,
$
x_{T^c,k}(w_{T^c\times \mathcal{I}}, y_{T^c}^{k-1})$,
$
\mathbf{x}_k(\mathbf{w}, \mathbf{y}^{k-1})$,
$
x_{T^c}^n(w_{{T^c}\times \mathcal{I}}, y_{T^c}^{n-1})$,  and
$
\mathbf{x}^n(\mathbf{w}, \mathbf{y}^{n-1}) $
 as done before~\eqref{convProofEq2}.
In addition, define the events
\begin{align*}
\mathcal{E}_{\mathbf{w}, y_i^n}&\stackrel{\mathrm{def}}{=} \left\{\hat w_{\mathcal{I}\times\{i\}} = w_{\mathcal{I}\times\{i\}}\right\},\\
\mathcal{G}_{\mathbf{w}, \mathbf{y}^n} &\stackrel{\mathrm{def}}{=} \left\{K^{[\mathbf{x}^n(\mathbf{w},\mathbf{y}^{n-1}), \mathbf{y}^n]} \in \mathcal{U}_{\boldsymbol{X}, \boldsymbol{Y}}^{(\delta^2, \mathbf{P}^{(\delta)})}\right\}\\
\noalign{\noindent and}
\mathcal{H}_{\mathbf{w}, \mathbf{y}^{n-1}} &\stackrel{\mathrm{def}}{=} \left\{\R[\mathbf{x}^n(\mathbf{w},\mathbf{y}^{n-1})] \in \mathcal{S}_{\delta P_{\min}}(\mathbf{P}^{(\delta)})\right\}
\end{align*}
to simplify notation. In order to simplify the RHS of~\eqref{AWGN:convProofEq1}, we write for each $\mathbf{w}\in \boldsymbol{\mathcal{W}}$
\begin{align}
&\Pr\left\{\left. \bigcap_{i\in \mathcal{I}} \left\{\varphi_i\left(w_{\{i\}\times\mathcal{I}},Y_i^n\right) = w_{\mathcal{I}\times\{i\}}\right\}\cap\left\{K^{[\boldsymbol{X}^n, \boldsymbol{Y}^n]}\in \mathcal{U}_{\boldsymbol{X}, \boldsymbol{Y}}^{(\delta^2, \mathbf{P}^{(\delta)})}\right\}\cap\left\{\R[\boldsymbol{X}^n] \in \mathcal{S}_{\delta P_{\min}}(\mathbf{P}^{(\delta)})\right\}\right|\boldsymbol{W}=\mathbf{w}\right\} \notag\\
& \quad= \int_{\mathbb{R}^{nN}} p_{\boldsymbol{Y}^n|\boldsymbol{W}=\mathbf{w}}(\mathbf{y}^n) \times \mathbf{1}\bigg\{\bigcap_{i\in \mathcal{I}} \mathcal{E}_{\mathbf{w}, y_i^n}\bigg\}  \mathbf{1}\left\{ \mathcal{G}_{\mathbf{w}, \mathbf{y}^n}\right\} \mathbf{1}\left\{\mathcal{H}_{\mathbf{w}, \mathbf{y}^{n-1}}\right\}\mathrm{d}\mathbf{y}^n \notag\\
&\quad \stackrel{\text{(a)}}{=}\int_{\mathbb{R}^{nN}} \prod_{k=1}^n  p_{\boldsymbol{Y}_k|\boldsymbol{X}_k}(\mathbf{y}_k|\mathbf{x}_k(\mathbf{w}, \mathbf{y}^{k-1}))\times \mathbf{1}\bigg\{\bigcap_{i\in \mathcal{I}} \mathcal{E}_{\mathbf{w}, y_i^n}\bigg\}  \mathbf{1}\left\{ \mathcal{G}_{\mathbf{w}, \mathbf{y}^n}\right\} \mathbf{1}\left\{\mathcal{H}_{\mathbf{w}, \mathbf{y}^{n-1}}\right\}\mathrm{d}\mathbf{y}^n \label{AWGN:convProofEq2}
\end{align}
where (a) follows from the fact due to Definitions~\ref{AWGN:defCode} and~\ref{AWGN:defDiscreteMemoryless} that
\begin{equation*}
 p_{\boldsymbol{Y}^n|\boldsymbol{W}=\mathbf{w}}(\mathbf{y}^n) = \prod_{k=1}^n  p_{\boldsymbol{Y}_k|\boldsymbol{X}_k}(\mathbf{y}_k|\mathbf{x}_k(\mathbf{w}, \mathbf{y}^{k-1}))
\end{equation*}
for all $\mathbf{y}^n\in \mathbb{R}^{nN}$.
\subsection{Further Simplifying the Correct Decoding Probability by Using the Method of Gaussian Types}
Define
\begin{align}
\gamma&\stackrel{\mathrm{def}}{=} \delta P_{\min}\label{AWGN:defGamma}
\\
\noalign{\noindent and}
\Delta&\stackrel{\mathrm{def}}{=} 1/n.\label{AWGN:defDelta}
\end{align}
 For each $\mathbf{w}\in\boldsymbol{\mathcal{W}}$ and each $(\Delta, \gamma, \mathbf{P}^{(\delta)})$-type $\mathbf{J}\in \mathcal{P}^{(\Delta, \gamma, \mathbf{P}^{(\delta)})}$, we define
\begin{equation}
\mathcal{A}^{(\Delta, \delta^2)}(\mathbf{w}; \mathbf{J})\stackrel{\mathrm{def}}{=} \left\{\mathbf{y}^n \in \mathbb{R}^{nN}\left|\left(\mathbf{x}^n(\mathbf{w},\mathbf{y}^{n-1}), \mathbf{y}^n\right)\in \mathcal{T}_\mathbf{J}^{(n,\Delta, \delta^2, \mathbf{P}^{(\delta)})}(\boldsymbol{X}, \boldsymbol{Y})\right.\right\} \label{AWGN:defSetA}
\end{equation}
and define for each non-empty $T\subsetneq\mathcal{I}$ and each $w_{T^c\times \mathcal{I}}\in \mathcal{W}_{T^c\times \mathcal{I}}$
\begin{equation}
\mathcal{F}_T^{(\Delta, \delta^2)}(w_{T^c\times \mathcal{I}};\mathbf{J})\stackrel{\mathrm{def}}{=} \left\{
y_{T^c}^n\in \mathbb{R}^{n|T^c|}\left|\,\text{$(x_{T^c}^n(w_{T^c\times\mathcal{I}}, y_{T^c}^n), y_{T^c}^n)\in \mathcal{T}_\mathbf{J}^{(n,\Delta, \delta^2, \mathbf{P}^{(\delta)})}(X_{T^c}, Y_{T^c})$}
\right.\right\}. \label{AWGN:defSetF}
\end{equation}
Since
\begin{align*}
&\bigcup\limits_{\mathbf{J}\in \mathcal{P}^{(\Delta, \gamma, \mathbf{P}^{(\delta)})}}\mathcal{T}_\mathbf{J}^{(n,\Delta, \delta^2, \mathbf{P}^{(\delta)})}(\boldsymbol{X}, \boldsymbol{Y}) \notag\\*
&\qquad\qquad\supseteq \left\{(\mathbf{x}^n(\mathbf{w},\mathbf{y}^{n-1}), \mathbf{y}^n)\in\boldsymbol{\mathcal{X}}\times\boldsymbol{\mathcal{Y}} \left|\,K^{[\mathbf{x}^n(\mathbf{w},\mathbf{y}^{n-1}), \mathbf{y}^n]} \in  \mathcal{U}_{\boldsymbol{X}, \boldsymbol{Y}}^{(\delta^2, \mathbf{P}^{(\delta)})}, \R{[\mathbf{x}^n(\mathbf{w},\mathbf{y}^{n-1})]}\in \mathcal{S}_\gamma(\mathbf{P}^{(\delta)}))\right.\right\}
\end{align*}
by~\eqref{AWGN:eqnTypeCovering*} and the definitions of $\mathcal{U}_{\boldsymbol{X}, \boldsymbol{Y}}^{(\delta^2, \mathbf{P}^{(\delta)})}$, $\mathcal{S}_{\gamma}(\mathbf{P}^{(\delta)})$, $\mathcal{T}_\mathbf{J}^{(n,\Delta, \delta^2, \mathbf{P}^{(\delta)})}(\boldsymbol{X}, \boldsymbol{Y})$ and $\mathcal{P}^{(\Delta, \gamma, \mathbf{P}^{(\delta)})}$ in~\eqref{AWGN:defSetU}, \eqref{AWGN:defSetSdelta}, Definition~\ref{AWGN:definitionJointTypeClass} and Definition~\ref{AWGN:definitionDeltaType} respectively, it together with the definition of~$\mathcal{A}^{(\Delta, \delta^2)}(\mathbf{w}; \mathbf{J})$ in~\eqref{AWGN:defSetA} implies that
$\bigcup\limits_{\mathbf{J}\in \mathcal{P}^{(\Delta, \gamma, \mathbf{P}^{(\delta)})}}\mathcal{A}^{(\Delta, \delta^2)}(\mathbf{w}; \mathbf{J})$ covers
\begin{align*}
 \left\{\mathbf{y}^n \in \mathbb{R}^{nN}\left| K^{[\mathbf{x}^n(\mathbf{w},\mathbf{y}^{n-1}), \mathbf{y}^n]} \in \mathcal{U}_{\boldsymbol{X}, \boldsymbol{Y}}^{(\delta^2, \mathbf{P}^{(\delta)})}, \R[\mathbf{x}^n(\mathbf{w},\mathbf{y}^{n-1})] \in \mathcal{S}_{\gamma}(\mathbf{P}^{(\delta)}) \right.\right\},
 \end{align*}
 which implies that the RHS of \eqref{AWGN:convProofEq2} can be bounded above as
 \begin{align}
& \int_{\mathbb{R}^{nN}} \prod_{k=1}^n  p_{\boldsymbol{Y}_k|\boldsymbol{X}_k}(\mathbf{y}_k|\mathbf{x}_k(\mathbf{w}, \mathbf{y}^{k-1})) \times \mathbf{1}\bigg\{\bigcap_{i\in \mathcal{I}} \mathcal{E}_{\mathbf{w}, y_i^n}\bigg\} \mathbf{1}\left\{ \mathcal{G}_{\mathbf{w}, \mathbf{y}^n}\right\} \mathbf{1}\left\{\mathcal{H}_{\mathbf{w}, \mathbf{y}^{n-1}}\right\}\mathrm{d}\mathbf{y}^n \notag\\
& \qquad \qquad\le \sum_{\mathbf{J}\in \mathcal{P}^{(\Delta, \gamma, \mathbf{P}^{(\delta)})}}\int\limits_{\substack{\mathcal{A}^{(\Delta, \delta^2)}(\mathbf{w}; \mathbf{J})}}\prod_{k=1}^n  p_{\boldsymbol{Y}_k|\boldsymbol{X}_k}(\mathbf{y}_k|\mathbf{x}_k(\mathbf{w}, \mathbf{y}^{k-1}))  \times  \mathbf{1}\bigg\{\bigcap_{i\in \mathcal{I}} \mathcal{E}_{\mathbf{w}, y_i^n}\bigg\}\mathrm{d}\mathbf{y}^n . \label{AWGN:convProofEq2*}
 \end{align}
 Combining~\eqref{AWGN:convProofEq1}, \eqref{AWGN:convProofEq2} and~\eqref{AWGN:convProofEq2*}, we conclude that
 the probability of correct decoding satisfies
\begin{align}
  1-\varepsilon_n
\le e^{-\tau n} + \frac{1}{|\boldsymbol{\mathcal{W}}|}\sum_{\mathbf{w}\in \boldsymbol{\mathcal{W}}} \sum_{\mathbf{J}\in \mathcal{P}^{(\Delta, \gamma, \mathbf{P}^{(\delta)})}}\int\limits_{\substack{ \mathcal{A}^{(\Delta, \delta^2)}(\mathbf{w}; \mathbf{J})}}\prod_{k=1}^n  p_{\boldsymbol{Y}_k|\boldsymbol{X}_k}(\mathbf{y}_k|\mathbf{x}_k(\mathbf{w}, \mathbf{y}^{k-1}))  \times \mathbf{1}\bigg\{\bigcap_{i\in \mathcal{I}} \mathcal{E}_{\mathbf{w}, y_i^n}\bigg\} \mathrm{d}\mathbf{y}^n . \label{AWGN:convProofEq3}
 \end{align}
  \subsection{Bounding the Correct Decoding Probability in Terms of $\mathcal{F}_T^{(\Delta, \delta^2)}(w_{T^c\times \mathcal{I}}; \mathbf{J})$}
Fix any arbitrary non-empty $T\subsetneq\mathcal{I}$. Define
  \begin{equation}
  a_T\stackrel{\mathrm{def}}{=}  \frac{1}{2}\log\left((2\pi e)^{|T^c|}|\Sigma_{T^c\times T^c}|\right)- \frac{\delta^2 N^3}{2\sigma_{\min}} \label{AWGN:defa}
  \end{equation}
  to simplify notation where $\sigma_{\min}>0$ is the smallest eigenvalue of~$\mathbf{\Sigma}$. In order to simplify the RHS of~\eqref{AWGN:convProofEq3}, we consider the innermost product therein. In particular, we consider the following chain of equalities for each $\mathbf{w}\in \boldsymbol{\mathcal{W}}$, each $\mathbf{J}\in\mathcal{P}^{(\Delta, \gamma, \mathbf{P}^{(\delta)})}$ and each $\mathbf{y}^n \in \mathcal{A}^{(\Delta, \delta^2)}(\mathbf{w}; \mathbf{J})$:
  \begin{align}
 \prod_{k=1}^n  p_{\boldsymbol{Y}_k|\boldsymbol{X}_k}(\mathbf{y}_k|\mathbf{x}_k(\mathbf{w}, \mathbf{y}^{k-1}))
 & = \prod_{k=1}^n  p_{Y_{T^c,k}|\boldsymbol{X}_k}(y_{T^c,k}|\mathbf{x}_k(\mathbf{w}, \mathbf{y}^{k-1}))  p_{Y_{T,k}|\boldsymbol{X}_k,Y_{T^c,k}}(y_{T,k}|\mathbf{x}_k(\mathbf{w}, \mathbf{y}^{k-1}), y_{T^c, k})\notag\\
 &\stackrel{\text{(b)}}{\le}  e^{-n a_T} \prod_{k=1}^n p_{Y_{T,k}|\boldsymbol{X}_k,Y_{T^c,k}}(y_{T,k}|\mathbf{x}_k(\mathbf{w}, \mathbf{y}^{k-1}), y_{T^c, k}) \label{AWGN:convProofEq3*}
  \end{align}
  where (b) follows from Lemma~\ref{AWGN:lemmaProductProbBound}.
 Following similar procedures for proving the chain of inequalities leading to~\eqref{convProofEq4}, we obtain the following inequality for each $\mathbf{J}\in\mathcal{P}^{(\Delta, \gamma, \mathbf{P}^{(\delta)})}$:
 \begin{align}
& \sum_{\mathbf{w}\in \boldsymbol{\mathcal{W}}}\int\limits_{\substack{\mathcal{A}^{(\Delta, \delta^2)}(\mathbf{w}; \mathbf{J})}}\prod_{k=1}^n  p_{\boldsymbol{Y}_k|\boldsymbol{X}_k}(\mathbf{y}_k|\mathbf{x}_k(\mathbf{w}, \mathbf{y}^{k-1})) \times \mathbf{1}\left\{\bigcap_{i\in \mathcal{I}} \mathcal{E}_{\mathbf{w}, y_i^n}\right\}\mathrm{d}\mathbf{y}^n  \notag\\
&\quad \stackrel{\eqref{AWGN:convProofEq3*}}{\le} e^{-n a_T} \sum_{\mathbf{w}\in \boldsymbol{\mathcal{W}}}\int\limits_{\substack{\mathcal{A}^{(\Delta, \delta^2)}(\mathbf{w}; \mathbf{J})}}          \prod_{k=1}^n p_{Y_{T,k}|\boldsymbol{X}_k,Y_{T^c,k}}(y_{T,k}|\mathbf{x}_k(\mathbf{w}, \mathbf{y}^{k-1}),y_{T^c,k}) \times \mathbf{1}\left\{\bigcap_{i\in T^c} \mathcal{E}_{\mathbf{w}, y_i^n}\right\}\mathrm{d}\mathbf{y}^n   \notag\\
 &\quad\le e^{-n a_T}  \sum_{w_{(T\times T^c)^c}\in \mathcal{W}_{(T\times T^c)^c}}\int\limits_{\substack{ \mathcal{F}_T^{(\Delta, \delta^2)}(w_{T^c\times \mathcal{I}}; \mathbf{J})}} 1\:  \mathrm{d}y_{T^c}^n. \label{AWGN:convProofEq4}
\end{align}
    \subsection{Bounding the Volume of $\mathcal{F}_T^{(\Delta, \delta^2)}(w_{T^c\times \mathcal{I}}; \mathbf{J})$}
  For each $\mathbf{J}=J_{\mathcal{I}\times\mathcal{I}}\in\mathcal{P}^{(\Delta, \gamma, \mathbf{P}^{(\delta)})}$, we let
\begin{equation*}
\phi_{X_{T^c}, Y_{T^c}}(x_{T^c}, y_{T^c}) \equiv \mathcal{N}\left(\left[\begin{matrix}x_{T^c}\\ y_{T^c}\end{matrix}\right]; 0^{2|T^c|},\left[\begin{matrix}J_{T^c\times T^c} & J_{T^c\times \mathcal{I}}G_{T^c\times \mathcal{I}}^t \\  G_{T^c\times \mathcal{I}} J_{\mathcal{I}\times T^c} & G_{T^c\times \mathcal{I}}  \mathbf{J}  G_{T^c\times \mathcal{I}}^t + \Sigma_{T^c\times T^c} \end{matrix}\right]\right)
 \end{equation*}
 denote the multivariate normal distribution in order to obtain an upper bound on the volume of $\mathcal{F}_T^{(\Delta, \delta^2)}(w_{T^c\times \mathcal{I}};\mathbf{J})$. For each $w_{T^c\times \mathcal{I}}\in \mathcal{W}_{T^c\times \mathcal{I}}$ and each $\mathbf{J}\in\mathcal{P}^{(\Delta, \gamma, \mathbf{P}^{(\delta)})}$, since the smallest eigenvalue of~$\mathbf{J}$ is at least $\gamma>0$ by the definition of $\mathcal{P}^{(\Delta, \gamma, \mathbf{P}^{(\delta)})}$ in Definition~\ref{AWGN:definitionDeltaType} and~\eqref{AWGN:defSetSdelta}, it follows from Proposition~\ref{AWGN:propositionInverseBound} that
\begin{align}
 \left(J_{T^c\times T^c}\right)^{-1}\in \Gamma_{\frac{N}{\gamma}}(0^{N\times N}), \label{AWGN:JTCInverseInProof}
\end{align}
and it is well known~\cite[Sec.~8.1.3]{IMM2012-03274} that
  \begin{align}
  \phi_{Y_{T^c}|X_{T^c}}(y_{T^c}|x_{T^c})&\equiv \mathcal{N}(y_{T^c}; \hat \mu_{T^c}(x_{T^c}; \mathbf{J}), \hat \Sigma_{T^c}(\mathbf{J}) ) \label{AWGN:defDistPhi}
\end{align}
where
\begin{align}
 \hat \mu_{T^c}(x_{T^c}; \mathbf{J})  &\stackrel{\mathrm{def}}{=}  G_{T^c\times \mathcal{I}} J_{\mathcal{I}\times T^c}  \left(J_{T^c\times T^c}\right)^{-1}x_{T^c} \label{AWGN:defDistPhiMean} \\
\noalign{\noindent  and}
 \hat \Sigma_{T^c}(\mathbf{J}) &\stackrel{\mathrm{def}}{=}  G_{T^c\times \mathcal{I}}  \mathbf{J}  G_{T^c\times \mathcal{I}}^t + \Sigma_{T^c\times T^c} - G_{T^c\times \mathcal{I}} J_{\mathcal{I}\times T^c}  \left(J_{T^c\times T^c}\right)^{-1} J_{T^c\times \mathcal{I}} G_{T^c\times \mathcal{I}}^t. \label{AWGN:defDistPhiCov}
 \end{align}
In the remainder of this subsection, we aim to show that the exponent of the volume of $\mathcal{F}_T^{(\Delta, \delta^2)}(w_{T^c\times \mathcal{I}}; \mathbf{J})$ is close to $\frac{1}{2}\log \left((2\pi e)^{|T^c|}| \hat \Sigma_{T^c}(\mathbf{J})|\right)$.
To this end, we write for each $w_{T^c\times \mathcal{I}}\in \mathcal{W}_{T^c\times \mathcal{I}}$ and each $\mathbf{J}\in\mathcal{P}^{(\Delta, \gamma, \mathbf{P}^{(\delta)})}$
\begin{align*}
\int_{\mathcal{F}_T^{(\Delta, \delta^2)}(w_{T^c\times \mathcal{I}}; \mathbf{J})} \prod_{k=1}^n \phi_{Y_{T^c}|X_{T^c}}(y_{T^c,k}|x_{T^c,k}(w_{T^c\times\mathcal{I}},y_{T^c}^{k-1}))\mathrm{d}y_{T^c}^n \le 1
\end{align*}
(recall the definition of~$\mathcal{F}_T^{(\Delta, \delta^2)}(w_{T^c\times \mathcal{I}}; \mathbf{J})$ in~\eqref{AWGN:defSetF}),
which implies by using~\eqref{AWGN:defDistPhi}, \eqref{AWGN:defDistPhiMean} and~\eqref{AWGN:defDistPhiCov} that
\begin{align}
\int_{\mathcal{F}_T^{(\Delta, \delta^2)}(w_{T^c\times \mathcal{I}}; \mathbf{J})}  e^{-n\left(\frac{1}{2}\log \left((2\pi)^{|T^c|}| \hat \Sigma_{T^c}(\mathbf{J})|\right)+\frac{1}{2n}\sum\limits_{k=1}^n\Tr\left((\hat \Sigma_{T^c}(\mathbf{J}))^{-1}\R{[y_{T^c,k}-\hat \mu_{T^c}(x_{T^c,k}(w_{T^c\times\mathcal{I}},y_{T^c}^{k-1});\mathbf{J})]}\right)\right)} \mathrm{d}y_{T^c}^n \le 1.  \label{AWGN:convProofEq4*}
\end{align}
Fix any $w_{T^c\times \mathcal{I}}\in \mathcal{W}_{T^c\times \mathcal{I}}$ and any $\mathbf{J}\in\mathcal{P}^{(\Delta, \gamma, \mathbf{P}^{(\delta)})}$. In order to obtain a lower bound on the LHS of~\eqref{AWGN:convProofEq4*}, we consider for each $y_{T^c}^n \in \mathcal{F}_T^{(\Delta, \delta^2)}(w_{T^c\times \mathcal{I}}; \mathbf{J})$
\begin{align}
\frac{1}{n}\sum_{k=1}^n \R{[y_{T^c,k}-\hat \mu_{T^c}(x_{T^c,k}(w_{T^c\times\mathcal{I}},y_{T^c}^{k-1});\mathbf{J})]}
& = \R{[y_{T^c}^n]} - \frac{2}{n}\sum_{k=1}^n \Upsilon^{\left[y_{T^c,k},G_{T^c\times \mathcal{I}}J_{\mathcal{I}\times T^c}(J_{T^c\times T^c})^{-1} x_{T^c,k}(w_{T^c\times\mathcal{I}}, y_{T^c}^{k-1})\right]}\notag\\
&\quad+ \frac{1}{n}\sum_{k=1}^n \R{[G_{T^c\times \mathcal{I}}J_{\mathcal{I}\times T^c}(J_{T^c\times T^c})^{-1} x_{T^c,k}(w_{T^c\times\mathcal{I}},y_{T^c}^{k-1})]}. \label{AWGN:convProofEq4*+}
\end{align}
Recalling the definitions of $\mathcal{F}_T^{(\Delta, \delta^2)}(w_{T^c\times \mathcal{I}}; \mathbf{J})$, $\mathcal{T}_{\mathbf{J}}^{(n, \Delta, \delta^2, \mathbf{P}^{(\delta)})}(X_{T^c}, Y_{T^c})$, $\mathcal{T}_{\mathbf{J}}^{(n, \Delta, \delta^2, \mathbf{P}^{(\delta)})}(\boldsymbol{X}, \boldsymbol{Y})$ and $\mathcal{U}_{\boldsymbol{X}, \boldsymbol{Y}}^{(\delta^2, \mathbf{P}^{(\delta)})}$ in~\eqref{AWGN:defSetF}, \eqref{AWGN:defJointTypeClassRestricted}, \eqref{AWGN:defJointTypeClass} and~\eqref{AWGN:defSetU} respectively, we conclude that for each $y_{T^c}^n \in \mathcal{F}_T^{(\Delta, \delta^2)}(w_{T^c\times \mathcal{I}}; \mathbf{J})$, there exists a pair $(\bar{\mathbf{x}}^n, \bar{\mathbf{y}}^n)\in \mathcal{T}_{\mathbf{J}}^{(n, \Delta, \delta^2, \mathbf{P}^{(\delta)})}(\boldsymbol{X}, \boldsymbol{Y})$ such that
\begin{align}
(\bar x_{T^c}^n,\bar y_{T^c}^n)&=( x_{T^c}^n(w_{T^c\times\mathcal{I}}, y_{T^c}^n), y_{T^c}^n), \label{AWGN:convProofEq4*++}
\\
\R{[\bar{\mathbf{x}}^n]}& \in \mathcal{V}^\Delta(\mathbf{J}) \label{AWGN:convProofEq4**}
\\
\noalign{\noindent and}
K^{[\bar{\mathbf{x}}^n, \bar{\mathbf{y}}^n]} &\in \mathcal{U}_{\boldsymbol{X}, \boldsymbol{Y}}^{(\delta^2, \mathbf{P}^{(\delta)})}. \label{AWGN:convProofEq4***}
\end{align}
It follows from~\eqref{AWGN:convProofEq4***}, the definition of~$\mathcal{U}_{\boldsymbol{X}, \boldsymbol{Y}}^{(\delta^2, \mathbf{P}^{(\delta)})}$ in~\eqref{AWGN:defSetU} and Proposition~\ref{AWGN:propositionProductBound} that
\begin{align}
\Upsilon^{[\bar{\mathbf{y}}^n,\bar{\mathbf{x}}^n]}&\in   \Gamma_{\delta^2}(\mathbf{G}\R{[\bar{\mathbf{x}}^n]}) \label{AWGN:convProofEq4*****}
\\
\noalign {\noindent and}
\R{[\bar{\mathbf{y}}^n]}&\in  \Gamma_{(2Ng_{\max}+1)\delta^2}(\mathbf{G}\R{[\bar{\mathbf{x}}^n]} \mathbf{G}^t+ \mathbf{\Sigma}) \label{AWGN:convProofEq4****}
\end{align}
whose derivations are provided in Appendix~\ref{AWGN:appendixE} for the sake of completeness.
Combining~\eqref{AWGN:convProofEq4**}, \eqref{AWGN:convProofEq4*****} and \eqref{AWGN:convProofEq4****}, we conclude that there exists a $Q^\Delta_{\mathcal{I}\times \mathcal{I}}\in \Gamma_{\Delta}(0^{N\times N})$ such that
\begin{align}
\R{[\bar x_{T^c}^n]} &= J_{T^c\times T^c} + Q^\Delta_{T^c\times T^c}, \label{AWGN:convProofEq4+}\\
\R{[\bar y_{T^c}^n]}&\in   \Gamma_{(2Ng_{\max}+1)\delta^2}(G_{T^c\times \mathcal{I}}(\mathbf{J}+Q^\Delta_{\mathcal{I}\times \mathcal{I}}) G_{T^c\times \mathcal{I}}^t+ \Sigma_{T^c\times T^c}), \label{AWGN:convProofEq4******}
\\
\noalign {\noindent and}
\Upsilon^{[\bar y_{T^c}^n, \bar x_{T^c}^n]}&\in   \Gamma_{\delta^2}(G_{T^c\times \mathcal{I}}(J_{\mathcal{I}\times T^c} + Q^\Delta_{\mathcal{I}\times T^c})). \label{AWGN:convProofEq4********}
\end{align}
Since
\begin{align*}
G_{T^c\times \mathcal{I}}Q^\Delta_{\mathcal{I}\times \mathcal{I}}&\in \Gamma_{N g_{\max} \Delta}(0^{T^c\times \mathcal{I}})
\\
\noalign{\noindent and}
G_{T^c\times \mathcal{I}}Q^\Delta_{\mathcal{I}\times \mathcal{I}} G_{T^c\times \mathcal{I}}^t &\in \Gamma_{N^2 g_{\max}^2 \Delta}(0^{T^c\times T^c})
\end{align*}
by Proposition~\ref{AWGN:propositionProductBound}, it follows from~\eqref{AWGN:convProofEq4******} and~\eqref{AWGN:convProofEq4********} that
\begin{align}
\Upsilon^{[\bar y_{T^c}^n, \bar x_{T^c}^n]}&\in\Gamma_{\delta^2 + N g_{\max} \Delta}(G_{T^c\times \mathcal{I}}J_{\mathcal{I}\times T^c}) \label{AWGN:convProofEq4********++}
\\
\noalign{\noindent and}
\R{[\bar y_{T^c}^n]}&\in \Gamma_{(2Ng_{\max}+1)\delta^2+ N^2 g_{\max}^2 \Delta}(G_{T^c\times \mathcal{I}}\mathbf{J} G_{T^c\times \mathcal{I}}^t+ \Sigma_{T^c\times T^c}). \label{AWGN:convProofEq4*******+}
\end{align}
Following~\eqref{AWGN:convProofEq4*+}, we use~\eqref{AWGN:JTCInverseInProof}, \eqref{AWGN:convProofEq4*++}, \eqref{AWGN:convProofEq4+}, \eqref{AWGN:convProofEq4********++}, \eqref{AWGN:convProofEq4*******+}, the fact that
$\mathbf{J}\in \Gamma_{(1+\delta)P_{\max}}(0^{N\times N})$ and Proposition~\ref{AWGN:propositionProductBound} to obtain
\begin{align}
\R{[y_{T^c}^n]}&\in \Gamma_{(2Ng_{\max}+1)\delta^2+ N^2 g_{\max}^2 \Delta}(G_{T^c\times \mathcal{I}}\mathbf{J} G_{T^c\times \mathcal{I}}^t+ \Sigma_{T^c\times T^c}), \label{AWGN:convProofEq4********+}\\
\noalign {}
&\hspace{-0.35 in}\frac{1}{n}\sum_{k=1}^n\Upsilon^{\left[y_{T^c,k},G_{T^c\times \mathcal{I}}J_{\mathcal{I}\times T^c}(J_{T^c\times T^c})^{-1} x_{T^c,k}(w_{T^c\times\mathcal{I}}, y_{T^c}^{k-1})\right]} \notag\\
&\in  \Gamma_{N^4 g_{\max}(\delta^2+ Ng_{\max} \Delta) (1+\delta)P_{\max}/\gamma}(G_{T^c\times \mathcal{I}}J_{\mathcal{I}\times T^c}(J_{T^c\times T^c})^{-1} J_{T^c\times \mathcal{I}}G_{T^c\times \mathcal{I}}^t)\label{AWGN:convProofEq4++2}
\\
\noalign {\noindent and}
&\hspace{-0.35 in}\frac{1}{n}\sum_{k=1}^n \R{[G_{T^c\times \mathcal{I}}J_{\mathcal{I}\times T^c}(J_{T^c\times T^c})^{-1} x_{T^c,k}(w_{T^c\times\mathcal{I}},y_{T^c}^{k-1})]}\notag\\
& \in \Gamma_{\Delta\left(N^4 g_{\max}  (1+\delta)P_{\max}/\gamma\right)^2}(G_{T^c\times \mathcal{I}}J_{\mathcal{I}\times T^c}(J_{T^c\times T^c})^{-1} J_{T^c\times \mathcal{I}}G_{T^c\times \mathcal{I}}^t)\label{AWGN:convProofEq4++3}.
\end{align}
To simplify notation, define
\begin{equation}
\kappa(\delta, \Delta)\stackrel{\mathrm{def}}{=}  (2Ng_{\max}+1)\delta^2+ N^2 g_{\max}^2 \Delta + 2N^4 g_{\max} (\delta^2+ Ng_{\max} \Delta) (1+\delta)P_{\max}/\gamma + \Delta\left(N^4 g_{\max}  (1+\delta)P_{\max}/\gamma\right)^2\!\!. \label{AWGN:defKappa}
\end{equation}
 Combining~\eqref{AWGN:convProofEq4*+}, \eqref{AWGN:convProofEq4********+}, \eqref{AWGN:convProofEq4++2}, \eqref{AWGN:convProofEq4++3} and~\eqref{AWGN:defKappa}, we obtain
 \begin{align*}
 &\frac{1}{n}\sum_{k=1}^n \R{[y_{T^c,k}-\hat \mu_{T^c}(x_{T^c,k}(w_{T^c\times\mathcal{I}},y_{T^c}^{k-1});\mathbf{J})]}\notag\\
 & \quad\in \Gamma_{\kappa(\delta, \Delta)}\left(G_{T^c\times \mathcal{I}}\mathbf{J} G_{T^c\times \mathcal{I}}^t + \Sigma_{T^c\times T^c} - G_{T^c\times \mathcal{I}}J_{\mathcal{I}\times T^c}(J_{T^c\times T^c})^{-1} J_{T^c\times \mathcal{I}}G_{T^c\times \mathcal{I}}^t\right),
 \end{align*}
 which implies by~\eqref{AWGN:defDistPhiCov} that
  \begin{align}
 \frac{1}{n}\sum_{k=1}^n \R{[y_{T^c,k}-\hat \mu_{T^c}(x_{T^c,k}(w_{T^c\times\mathcal{I}},y_{T^c}^{k-1});\mathbf{J})]} \in \Gamma_{\kappa(\delta, \Delta)}\left(\hat \Sigma_{T^c}(\mathbf{J})\right). \label{AWGN:convProofEq4++4}
 \end{align}
 Since
 \begin{align}
  \hat \Sigma_{T^c}(\mathbf{J})= G_{T^c\times T} \left( J_{T\times T} - J_{T\times T^c}  \left(J_{T^c\times T^c}\right)^{-1} J_{T^c\times T} \right) G_{T^c\times T}^t  + \Sigma_{T^c\times T^c} \label{AWGN:defHatSigma*}
 \end{align}
 by simplifying~\eqref{AWGN:defDistPhiCov}, it follows that all the eigenvalues of $\hat \Sigma_{T^c}(\mathbf{J})$ are at least $\sigma_{\min}$, which implies by Proposition~\ref{AWGN:propositionInverseBound} that
 \begin{align}
 (\hat \Sigma_{T^c}(\mathbf{J}))^{-1} \in \Gamma_{\frac{N}{\sigma_{\min}}}(0^{T^c\times T^c}). \label{AWGN:convProofEq4++5}
 \end{align}
 Therefore,
 it follows from~\eqref{AWGN:convProofEq4++4}, \eqref{AWGN:convProofEq4++5} and Proposition~\ref{AWGN:propositionInverseBound} that
 \begin{align*}
 \left|\frac{1}{n}\sum_{k=1}^n\Tr\left((\hat \Sigma_{T^c}(\mathbf{J}))^{-1}\R{[y_{T^c,k}-\hat \mu_{T^c}(x_{T^c,k}(w_{T^c\times\mathcal{I}},y_{T^c}^{k-1});\mathbf{J})]}\right)\right|\le \frac{N^2\kappa(\delta, \Delta)}{\sigma_{\min}},
 \end{align*}
which together with~\eqref{AWGN:convProofEq4*} implies that
 \begin{align}
\int_{\mathcal{F}_T^{(\Delta, \delta^2)}(w_{T^c\times \mathcal{I}}; \mathbf{J})}1\:\mathrm{d}y_{T^c}^n
\le e^{n\left(\frac{1}{2}\log \left((2\pi e)^{|T^c|}| \hat \Sigma_{T^c}(\mathbf{J})|\right)+ \frac{N^2\kappa(\delta, \Delta)}{2\sigma_{\min}}\right)}. \label{AWGN:convProofEq5}
\end{align}

\subsubsection{Showing the Exponential Decay of the Probability of Correct Decoding}
Combining~\eqref{AWGN:convProofEq4}, \eqref{AWGN:defa} and~\eqref{AWGN:convProofEq5} and using the fact due to~\eqref{defAlphabet} that
\begin{equation*}
\frac{|\mathcal{W}_{(T\times T^c)^c}|}{|\boldsymbol{\mathcal{W}}|} = \frac{1}{\prod\limits_{(i,j)\in T\times T^c}\lceil e^{n (1-\delta)R_{i,j}}\rceil} \le e^{-n\sum\limits_{(i,j)\in T\times T^c}(1-\delta)R_{i,j}},
\end{equation*}
 we have for each $\mathbf{J}\in\mathcal{P}^{(\Delta, \gamma, \mathbf{P}^{(\delta)})}$
\begin{align}
 &\frac{1}{|\boldsymbol{\mathcal{W}}|}\sum_{\mathbf{w}\in \boldsymbol{\mathcal{W}}}\int\limits_{\substack{\mathcal{A}^{(\Delta, \delta^2)}(\mathbf{w}; \mathbf{J})}}\prod_{k=1}^n  p_{\boldsymbol{Y}_k|\boldsymbol{X}_k}(\mathbf{y}_k|\mathbf{x}_k(\mathbf{w}, \mathbf{y}^{k-1})) \times \mathbf{1}\left\{\bigcap_{i\in \mathcal{I}} \mathcal{E}_{\mathbf{w}, y_i^n}\right\}\mathrm{d}\mathbf{y}^n  \notag\\
 &\quad \le e^{-n\big(\sum\limits_{(i,j)\in T\times T^c}(1-\delta)R_{i,j}- \frac{1}{2}\log\left(\frac{|\hat \Sigma_{T^c}(\mathbf{J})|}{|\Sigma_{T^c\times T^c}|}\right) - \frac{\delta^2 N^3 + N^2\kappa(\delta, \Delta)}{2\sigma_{\min}} \big)}. \label{AWGN:convProofEq6}
\end{align}
Using~\eqref{AWGN:defEta}, \eqref{AWGN:sufficientlyLargeNInProof}, \eqref{AWGN:defGamma}, \eqref{AWGN:defDelta} and~\eqref{AWGN:defKappa}, we have
\begin{align}
\frac{\delta^2 N^3 +N^2\kappa(\delta, \Delta)}{2\sigma_{\min}} \le \eta(\delta). \label{AWGN:convProofEq7}
\end{align}
Combining~\eqref{AWGN:defHatSigma*}, \eqref{AWGN:convProofEq6} and~\eqref{AWGN:convProofEq7}, we have for each $\mathbf{J}\in\mathcal{P}^{(\Delta, \gamma, \mathbf{P}^{(\delta)})}$
\begin{align}
 &\frac{1}{|\boldsymbol{\mathcal{W}}|}\sum_{\mathbf{w}\in \boldsymbol{\mathcal{W}}}\int\limits_{\substack{\mathcal{A}^{(\Delta, \delta^2)}(\mathbf{w}; \mathbf{J})}}\prod_{k=1}^n  p_{\boldsymbol{Y}_k|\boldsymbol{X}_k}(\mathbf{y}_k|\mathbf{x}_k(\mathbf{w}, \mathbf{y}^{k-1})) \times \mathbf{1}\left\{\bigcap_{i\in \mathcal{I}} \mathcal{E}_{\mathbf{w}, y_i^n}\right\}\mathrm{d}\mathbf{y}^n  \notag\\
\notag\\*
&\qquad \le e^{-n\big(\sum\limits_{(i,j)\in T\times T^c}(1-\delta)R_{i,j}- \frac{1}{2}\log\left|I_{|T^c|} + G_{T^c\times T}J_{T|T^c}G_{T^c\times T}^t \left(\Sigma_{T^c\times T^c}\right)^{-1}\right| - \eta(\delta) \big)} \label{AWGN:convProofEq8}
\end{align}
for any non-empty $T\subsetneq\mathcal{I}$ where $J_{T|T^c}\stackrel{\mathrm{def}}{=}J_{T\times T} - J_{T\times T^c}  \left(J_{T^c\times T^c}\right)^{-1} J_{T^c\times T}$.
Using~\eqref{AWGN:convProofEq3}, \eqref{AWGN:convProofEq8}, \eqref{AWGN:convProofRate*}, \eqref{AWGN:closedFormK}, Corollary~\ref{AWGN:corollaryCardinalityBound} and~\eqref{AWGN:defDelta}, we obtain
 \begin{align}
  1-\varepsilon_n
 &\le e^{-\tau n} + \prod\limits_{(i,j)\in \mathcal{I}\times \mathcal{I}}\left(2\left\lceil n\sqrt{P_i P_j}\right\rceil+1\right)e^{-n \eta(\delta) }\notag \\
 & \le  e^{-\tau n} + (2nP_{\max}+3)^{N^2}e^{-n \eta(\delta) }. \label{AWGN:convProofEq9}
\end{align}
Consequently, \eqref{AWGN:convProofError} holds by~\eqref{AWGN:convProofEq9} as $\eta(\delta)$ is positive by~\eqref{AWGN:defEta} and hence~\eqref{AWGN:convProofSt} holds for all $\varepsilon\in[0,1)$.

\section{Concluding Remarks} \label{conclusion}
This paper presents the first complete proof of the strong converse theorem for any DMN with tight cut-set bound. The proof is based on the method of types.
In addition, the strong converse theorem is generalized to any Gaussian network with tight cut-set bound under almost-sure power constraints.
Our generalization of the strong converse proof for DMNs to Gaussian networks is not obvious, mainly due to the fact that the strong converse proof for DMNs is based on the method of types~\cite[Ch.~2]{Csi97}.
More specifically, the method of types defined for DMNs is based on counting arguments since the input and output alphabets of DMNs are finite. On the contrary, the method of types defined for Gaussian networks is based on careful approximation and quantization arguments due to the continuous input and output alphabets.
There is one key difference between the proof for DMNs in Section~\ref{secConverse} and the proof for Gaussian networks in Section~\ref{AWGN:secConverse}:
In the proof for Gaussian networks, we avoid using conditional types, which cannot be easily defined when the correlation between the input symbols and the noise random variables is not negligible. Instead, we define joint type classes in a novel way in Definition~\ref{AWGN:definitionJointTypeClass} so that we can omit the use of conditional types in our proof. In contrast, the proof for DMNs in Section~\ref{secConverse} relies heavily on the definition of conditional types.

Important consequences of the two strong converse theorems are new strong converses for the Gaussian MAC with feedback and the following relay channels under both the discrete memoryless and the Gaussian models: The degraded RC, the RC with orthogonal sender components, and the general RC with feedback. The strong converse theorem for the Gaussian case complements the following recent findings: If long-term power constraints are used instead of almost-sure power constraints, then the strong converse does not hold for the Gaussian degraded RC~\cite{FongTan16GaussianRelay} and the Gaussian MAC with feedback~\cite{LFT16MACfeedback}.

\section{Appendix}

\subsection{Proof of Lemma~\ref{AWGN:lemmaJointTypeExponentiallySmall}}\label{AWGN:appendixB}
Before proving Lemma~\ref{AWGN:lemmaJointTypeExponentiallySmall}, we need to prove two preliminary results. The following proposition states that the probability that the empirical autocorrelation of $\boldsymbol{Z}^n$ falls outside $\Gamma_\delta(\mathbf{\Sigma})$ is exponentially small. The proof of Proposition~\ref{AWGN:propositionExponentiallySmall} is due to the theory of large deviations~\cite{Sanov} and is provided here for the sake of completeness.
\smallskip
\begin{Proposition} \label{AWGN:propositionExponentiallySmall}
Let $p_{\boldsymbol{Z}}(\mathbf{z})=\mathcal{N}(\mathbf{z}; 0^N, \mathbf{\Sigma})$ for all $\mathbf{z}$, let $\boldsymbol{Z}^n$ be $n$ independent copies of $\boldsymbol{Z}\sim p_{\boldsymbol{Z}}$, and let $p_{\boldsymbol{Z}^n}$ be the distribution of $\boldsymbol{Z}^n$, i.e.,
\begin{equation*}
p_{\boldsymbol{Z}^n}(\mathbf{z}^n) = \prod_{k=1}^n p_{\boldsymbol{Z}}(\mathbf{z}_k)
\end{equation*}
for all $\mathbf{z}^n$.
For any $\delta>0$, there exists a constant $\tau>0$ which is a function of $\mathbf{\Sigma}$ such that for all sufficiently large~$n$,
\begin{align}
\int_{\mathbb{R}^{nN}}p_{\boldsymbol{Z}^n}(\mathbf{z}^n)\times \mathbf{1}\left\{\R{[\mathbf{z}^n]}\in \Gamma_\delta(\mathbf{\Sigma})\right\}\mathrm{d}\mathbf{z}^n > 1-e^{-\tau n}. \label{AWGN:typeZnProbSmall}
\end{align}
\end{Proposition}
\begin{proof}
Let $t>0$ be any real number.
Consider the following chain of inequalities for each $(i,j)\in\mathcal{I}\times \mathcal{I}$, where all the probability and expectation terms are evaluated according to $p_{\boldsymbol{Z}^n}$:
\begin{align}
\Pr\left\{\left|\frac{1}{n}\sum_{k=1}^n Z_{i,k}Z_{j,k} - \E[Z_{i,k}Z_{j,k}]\right| > \delta\right\}
& = 2\,\Pr\left\{\frac{1}{n}\sum_{k=1}^n Z_{i,k}Z_{j,k}  > \E[Z_{i,k}Z_{j,k}]+\delta\right\}\notag\\
& \stackrel{\text{(a)}}{\le}\frac{2\left(\E[e^{t Z_{i,k}Z_{j,k}}]\right)^n}{e^{tn(\E[Z_{i,k}Z_{j,k}]+\delta)}}\notag\\
& = 2e^{tn\left(\frac{1}{t}\log\E[e^{t Z_{i,k}Z_{j,k}}] -\E[Z_{i,k}Z_{j,k}]-\delta\right)} \label{AWGN:eq3ProofofLemmaExponential}
\end{align}
where (a) follows from Chernoff's bound. Since
\begin{align*}
\lim_{t\rightarrow 0} \frac{1}{t}\log\E[e^{t Z_{i,k}Z_{j,k}}] = \E[Z_{i,k}Z_{j,k}],
\end{align*}
there exists a sufficiently small $t_{ij}>0$ which is a function of~$\mathbf{\Sigma}$ such that
\begin{align*}
\frac{1}{t_{ij}}\log\E[e^{t_{ij} Z_{i,k}Z_{j,k}}] - \E[Z_{i,k}Z_{j,k}] \le \delta/2,
\end{align*}
which implies by using~\eqref{AWGN:eq3ProofofLemmaExponential} that
\begin{align}
\Pr\left\{\left|\frac{1}{n}\sum_{k=1}^n Z_{i,k}Z_{j,k} - \E[Z_{i,k}Z_{j,k}]\right| > \delta\right\} \le 2e^{-t_{ij}\delta n/2}. \label{AWGN:eq4ProofofLemmaExponential}
\end{align}
Since there exists a finite set of positive numbers $\{t_{ij}>0\,|\,(i,j)\in\mathcal{I}\times \mathcal{I}\}$ such that~\eqref{AWGN:eq4ProofofLemmaExponential} holds for all $(i,j)\in \mathcal{I}$, we conclude that~\eqref{AWGN:typeZnProbSmall} holds for all sufficiently large~$n$ by choosing $\tau\stackrel{\mathrm{def}}{=} \frac{\delta}{4}\min_{(i,j)\in \mathcal{I}\times \mathcal{I}}\{t_{ij}\}$.
\end{proof}
\smallskip
The following proposition states that the probability that the empirical correlation between $\boldsymbol{X}^n$ and $\boldsymbol{Z}^n$ falls outside $\Gamma_\delta(0^{N\times N})$ is exponentially small. The proof of Proposition~\ref{AWGN:propositionCorrelationExponentiallySmall} is based on Chernoff's bound and the almost-sure power constraints~\eqref{AWGN:peakPowerConstraints}.
\smallskip
\begin{Proposition}  \label{AWGN:propositionCorrelationExponentiallySmall}
For any $\delta>0$, there exists a constant $\tau>0$ which is a function of $(\mathbf{P},\mathbf{\Sigma})$ such that for all sufficiently large~$n$,
\begin{align}
\int_{\mathbb{R}^{nN}}\int_{\mathbb{R}^{nN}}p_{\boldsymbol{X}^n,\boldsymbol{Z}^n}(\mathbf{x}^n, \mathbf{z}^n)\times \mathbf{1}\left\{\Upsilon^{[\mathbf{x}^n, \mathbf{z}^n]}\in \Gamma_\delta(0^{N\times N})\right\}\mathrm{d}\mathbf{z}^n \mathrm{d}\mathbf{x}^n > 1-e^{-\tau n} \label{AWGN:typeXnZnProbSmall}
\end{align}
holds for any $(n, \mathbf{R}, \mathbf{P})$-code where $p_{\boldsymbol{X}^n,\boldsymbol{Z}^n}$ is the distribution induced by the code.
\end{Proposition}
\begin{proof}
Let $t>0$ be any real number. Let $\sigma_j^2>0$ be the variance of $Z_{j,k}$ for each $j\in\mathcal{I}$ and each~$k\in\{1, 2, \ldots, n\}$. Fix a $\delta>0$ and any $(n, \mathbf{R}, \mathbf{P})$-code. Consider the following chain of inequalities for each $(i,j)\in\mathcal{I}\times \mathcal{I}$, where all the probability and expectation terms are evaluated according to the distribution induced by the $(n, \mathbf{R}, \mathbf{P})$-code (cf.\ \eqref{AWGN:inputOutputRelation}):
\begin{align}
\Pr\left\{\left|\frac{1}{n}\sum_{k=1}^n X_{i,k}Z_{j,k}\right| > \delta\right\}
& = 2\,\Pr\left\{\frac{1}{n}\sum_{k=1}^n X_{i,k}Z_{j,k}  > \delta\right\}\notag\\
& \stackrel{\text{(a)}}{\le}\frac{2\E\left[e^{t\sum_{k=1}^n X_{i,k}Z_{j,k}}\right]}{e^{\delta t n}}\notag\\
& \stackrel{\text{(b)}}{\le} \frac{2}{e^{\delta t n}}\, \E\bigg[e^{t\sum_{k=1}^n X_{i,k}Z_{j,k} + \frac{t^2 \sigma_j^2}{2}\sum_{k=1}^n (P_i-X_{i,k}^2)}\bigg] \label{AWGN:eq1ProofofLemmaCorrelationExponential}
\end{align}
where
\begin{enumerate}
\item[(a)] follows from Chernoff's bound.
 \item[(b)] follows from the almost-sure power constraint~\eqref{AWGN:peakPowerConstraints} for node~$i$.
 \end{enumerate}
 Since $Z_{j,k}$ is independent of $(X_i^k,Z_j^{k-1})$ for each $k\in\{1, 2, \ldots, n\}$, straightforward calculations reveal that
\begin{align}
\E\left[e^{t\sum_{k=1}^n X_{i,k}Z_{j,k} - \frac{t^2 \sigma_j^2}{2}\sum_{k=1}^n X_{i,k}^2}\right] = 1. \label{AWGN:eq2ProofofLemmaCorrelationExponential}
\end{align}
Combining~\eqref{AWGN:eq1ProofofLemmaCorrelationExponential} and~\eqref{AWGN:eq2ProofofLemmaCorrelationExponential}, we have for each $(i,j)\in \mathcal{I}\times \mathcal{I}$
\begin{align*}
\Pr\left\{\left|\frac{1}{n}\sum_{k=1}^n X_{i,k}Z_{j,k}\right| > \delta\right\} &\le \frac{2 e^{\sigma_j^2 t_{ij}^2n P_i/2 }}{e^{\delta t_{ij} n}}\\
\noalign{\noindent for any $t_{ij}>0$, which implies by choosing $t_{ij}\stackrel{\mathrm{def}}{=} \frac{\delta}{\sigma_j^2 P_i}$ that}
\Pr\left\{\left|\frac{1}{n}\sum_{k=1}^n X_{i,k}Z_{j,k}\right| > \delta\right\} &\le 2 e^{- \frac{\delta^2 n}{2\sigma_j^2P_i}}
\end{align*}
for all $(i,j)\in \mathcal{I}\times \mathcal{I}$. Consequently, \eqref{AWGN:typeXnZnProbSmall} holds for all sufficiently large~$n$ by choosing $\tau\stackrel{\mathrm{def}}{=} \frac{\delta^2}{4 \max\limits_{(i,j)\in \mathcal{I}\times \mathcal{I}}\sigma_j^2 P_i}$.
\end{proof}
\smallskip

We are ready to present the proof of Lemma~\ref{AWGN:lemmaJointTypeExponentiallySmall}.
Fix a $\delta>0$ and any $(n, \mathbf{R}, \mathbf{P})$-code. Let $p_{\boldsymbol{X}^n, \boldsymbol{Z}^n, \boldsymbol{Y}^n}$ be the distribution induced by the code (cf.\ \eqref{AWGN:inputOutputRelation}). Using the union bound, \eqref{AWGN:typeXnProb1}, the definition of $K^{[\mathbf{x}^n,\mathbf{y}^n]}$ in~\eqref{AWGN:defKCov} and the definition of $\mathcal{U}_{\boldsymbol{X}, \boldsymbol{Y}}^{(\delta, \mathbf{P})}$ in~\eqref{AWGN:defSetU}, we have for all sufficiently large~$n$
\begin{align}
&\int_{\mathbb{R}^{nN}}\int_{\mathbb{R}^{nN}}p_{\boldsymbol{X}^n, \boldsymbol{Y}^n}(\mathbf{x}^n,\mathbf{y}^n)\times \mathbf{1}\left\{K^{[\mathbf{x}^n,\mathbf{y}^n]}\notin \mathcal{U}_{\boldsymbol{X}, \boldsymbol{Y}}^{(\delta, \mathbf{P})}\right\}\mathrm{d}\mathbf{y}^n\mathrm{d}\mathbf{x}^n \notag\\
& \le  \int_{\mathbb{R}^{nN}}\int_{\mathbb{R}^{nN}}p_{\boldsymbol{X}^n, \boldsymbol{Y}^n}(\mathbf{x}^n,\mathbf{y}^n)\times\mathbf{1}\left\{\Upsilon^{[\mathbf{x}^n, \mathbf{y}^n]}- \R{[\mathbf{x}^n]}\mathbf{G}^t\notin \Gamma_\delta(0^{N\times N})\right\}\mathrm{d}\mathbf{y}^n\mathrm{d}\mathbf{x}^n \notag\\
&\quad + \int_{\mathbb{R}^{nN}}\int_{\mathbb{R}^{nN}}p_{\boldsymbol{X}^n, \boldsymbol{Y}^n}(\mathbf{x}^n,\mathbf{y}^n) \times \mathbf{1}\left\{\Upsilon^{[\mathbf{y}^n, \mathbf{x}^n]}-\mathbf{G}\R{[\mathbf{x}^n]}\notin \Gamma_\delta(0^{N\times N})\right\}\mathrm{d}\mathbf{y}^n\mathrm{d}\mathbf{x}^n\notag\\
&\quad + \int_{\mathbb{R}^{nN}}\int_{\mathbb{R}^{nN}}p_{\boldsymbol{X}^n, \boldsymbol{Y}^n}(\mathbf{x}^n,\mathbf{y}^n) \times \mathbf{1}\left\{\R{[\mathbf{y}^n]} + \mathbf{G}\R{[\mathbf{x}^n]}\mathbf{G}^t - \mathbf{G}\Upsilon^{[\mathbf{x}^n, \mathbf{y}^n]}- \Upsilon^{[\mathbf{y}^n, \mathbf{x}^n]}\mathbf{G}^t\notin \Gamma_\delta(\Sigma)\right\}\mathrm{d}\mathbf{y}^n\mathrm{d}\mathbf{x}^n \label{AWGN:appendixBeq1}.
\end{align}
Using Definition~\ref{AWGN:defCorrelationMatrix} and letting $\mathbf{z}^n \stackrel{\mathrm{def}}{=}  \mathbf{y}^n - \mathbf{G} \mathbf{x}^n$, we have
\begin{align}
\Upsilon^{[\mathbf{x}^n, \mathbf{y}^n]}- \R{[\mathbf{x}^n]}\mathbf{G}^t & = \Upsilon^{[\mathbf{x}^n, \mathbf{z}^n]}, \label{AWGN:appendixBeq2}\\
 \Upsilon^{[\mathbf{y}^n, \mathbf{x}^n]}-\mathbf{G}\R{[\mathbf{x}^n]} &= \Upsilon^{[\mathbf{z}^n, \mathbf{x}^n]} \label{AWGN:appendixBeq3}
\\
\noalign{\noindent and}
\R{[\mathbf{y}^n]} + \mathbf{G}\R{[\mathbf{x}^n]}\mathbf{G}^t - \mathbf{G}\Upsilon^{[\mathbf{x}^n, \mathbf{y}^n]}- \Upsilon^{[\mathbf{y}^n, \mathbf{x}^n]}\mathbf{G}^t &= \R{[\mathbf{z}^n]}. \label{AWGN:appendixBeq4}
\end{align}
Combining the channel law~\eqref{AWGN:inputOutputRelation}, \eqref{AWGN:appendixBeq1}, \eqref{AWGN:appendixBeq2}, \eqref{AWGN:appendixBeq3}, \eqref{AWGN:appendixBeq4} and applying Proposition~\ref{AWGN:propositionExponentiallySmall} and Proposition~\ref{AWGN:propositionCorrelationExponentiallySmall}, we have
\begin{align*}
\int_{\mathbb{R}^{nN}}\int_{\mathbb{R}^{nN}}p_{\boldsymbol{X}^n, \boldsymbol{Y}^n}(\mathbf{x}^n,\mathbf{y}^n)\times \mathbf{1}\left\{K^{[\mathbf{x}^n,\mathbf{y}^n]}\notin \mathcal{U}_{\boldsymbol{X}, \boldsymbol{Y}}^{(\delta, \mathbf{P})}\right\}\mathrm{d}\mathbf{y}^n\mathrm{d}\mathbf{x}^n \le e^{-\lambda n}
\end{align*}
for some $\lambda>0$ that depends on $\mathbf{P}$ and $\mathbf{\Sigma}$. This completes the proof.

    \subsection{Proof of Proposition~\ref{AWGN:propositionCardinalityBound}}\label{AWGN:appendixC}
  Fix a $\Delta>0$ and a $\gamma>0$.  Since~$\mathcal{S}(\mathbf{P})$ defined in~\eqref{AWGN:defSetS} is a set of covariance matrices, it follows that $\mathcal{S}(\mathbf{P})$ is a bounded set that is contained in
\begin{equation*}
\bar{\mathcal{S}}(\mathbf{P}) = \left\{\mathbf{K}\in \mathbb{R}^{N\times N}\left|\,
\parbox[c]{3.5 in}{$\mathbf{K}\succeq 0$ where the $ij^{\text{th}}$ element $k_{ij}$ satisfies $|k_{ij}|\le \sqrt{P_i P_j}$ for all $(i,j)\in\mathcal{I}\times \mathcal{I}$}
\right.\right\}. 
\end{equation*}
Define
\begin{equation*}
\bar{\mathcal{S}}^\Delta(\mathbf{P}) \stackrel{\mathrm{def}}{=} \left\{\mathbf{K}\in \mathbb{R}^{N\times N}\left|\,
\parbox[c]{3.75 in}{$\mathbf{K}\succeq 0$ where the $ij^{\text{th}}$ element $k_{ij}$ satisfies $|k_{ij}|\le \left\lceil\frac{\sqrt{P_i P_j}}{\Delta}\right\rceil \Delta$ for all $(i,j)\in\mathcal{I}\times \mathcal{I}$}
\right.\right\}.
\end{equation*}
Since $\mathcal{S}_\gamma(\mathbf{P}) \subseteq \mathcal{S}(\mathbf{P}) \subseteq \bar{\mathcal{S}}(\mathbf{P}) \subseteq \bar{\mathcal{S}}^\Delta(\mathbf{P})$ and $\bar{\mathcal{S}}^\Delta(\mathbf{P})$ contains at most $\prod\limits_{(i,j)\in \mathcal{I}\times \mathcal{I}}\left(2\left\lceil\frac{\sqrt{P_i P_j}}{\Delta}\right\rceil+1\right)$ $\Delta$-quantizers (cf.\ Definition~\ref{AWGN:definitionTypeRep}),
it follows that $\mathcal{S}_\gamma(\mathbf{P})$ contains at most $\prod\limits_{(i,j)\in \mathcal{I}\times \mathcal{I}}\left(2\left\lceil\frac{\sqrt{P_i P_j}}{\Delta}\right\rceil+1\right)$ $\Delta$-quantizers, which together with the definition of $\mathcal{L}^{(\Delta, \gamma, \mathbf{P})}$ in~\eqref{AWGN:defDeltaInputType} implies that
\begin{equation*}
|\mathcal{L}^{(\Delta, \gamma, \mathbf{P})}| \le \prod\limits_{(i,j)\in \mathcal{I}\times \mathcal{I}}\left(2\left\lceil\frac{\sqrt{P_i P_j}}{\Delta}\right\rceil+1\right).
\end{equation*}

\subsection{Proof of Lemma~\ref{AWGN:lemmaProductProbBound}} \label{AWGN:appendixD}
For each $(\mathbf{x}^n, \mathbf{y}^n)\in \mathcal{T}_{\mathbf{J}}^{(n, \Delta, \delta, \mathbf{P})}(\boldsymbol{X}, \boldsymbol{Y})$, consider
\begin{align}
\prod_{k=1}^n q_{Y_{T^c}|\boldsymbol{X}}(y_{T^c,k}|\mathbf{x}_k(\mathbf{w}, \mathbf{y}^{k-1}))
& \stackrel{\text{(a)}}{=} \prod_{k=1}^n \mathcal{N}(y_{T^c,k}; G_{T^c\times \mathcal{I}}\, \mathbf{x}_k(\mathbf{w}, \mathbf{y}^{k-1}), \Sigma_{T^c\times T^c})\notag\\
&\stackrel{\eqref{AWGN:normalDist}}{=}
 e^{-n\left(\frac{1}{2}\log \left((2\pi)^{|T^c|} |\Sigma_{T^c\times T^c}|\right) + \frac{1}{2n}\sum_{k=1}^n \Tr\left( \left(\Sigma_{T^c\times T^c}\right)^{-1}\R{[y_{T^c,k}-G_{T^c\times \mathcal{I}}\, \mathbf{x}_k(\mathbf{w}, \mathbf{y}^{k-1})]}\right)\right)}\notag\\
 &=  e^{-n\left(\frac{1}{2}\log \left((2\pi)^{|T^c|} |\Sigma_{T^c\times T^c}|\right) + \frac{1}{2} \Tr\left( \left(\Sigma_{T^c\times T^c}\right)^{-1}\frac{1}{n}\sum_{k=1}^n\R{[y_{T^c,k}-G_{T^c\times \mathcal{I}}\, \mathbf{x}_k(\mathbf{w}, \mathbf{y}^{k-1})]}\right)\right)} \label{AWGN:eq1ProofofLemmaProduct}
\end{align}
where (a) is due to Definition~\ref{AWGN:defDiscreteMemoryless}.
 By the definitions of $\mathcal{T}_{\mathbf{J}}^{(n, \Delta, \delta, \mathbf{P})}(\boldsymbol{X}, \boldsymbol{Y})$ and~$\mathcal{U}_{\boldsymbol{X}, \boldsymbol{Y}}^{(\delta, \mathbf{P})}$ in~\eqref{AWGN:defJointTypeClass} and~\eqref{AWGN:defSetU} respectively, we have
   \begin{align}
 \R{[\mathbf{y}^n]}+ \mathbf{G} \R{[\mathbf{x}^n]}\mathbf{G}^t - \mathbf{G} \Upsilon^{[\mathbf{x}^n, \mathbf{y}^n]}-  \Upsilon^{[\mathbf{y}^n, \mathbf{x}^n]}\mathbf{G}^t &\in \Gamma_\delta(\mathbf{\Sigma}), \notag
\\
 \noalign{\noindent which implies that}
\frac{1}{n}\sum_{k=1}^n \R{[\mathbf{y}_k-\mathbf{G} \mathbf{x}_k(\mathbf{w}, \mathbf{y}^{k-1})]} &\in \Gamma_{\delta}(\mathbf{\Sigma}), \notag
\\
\noalign{\noindent which then implies that}
\frac{1}{n}\sum_{k=1}^n\R{[y_{T^c,k}-G_{T^c\times \mathcal{I}}\, \mathbf{x}_k(\mathbf{w}, \mathbf{y}^{k-1})]} &\in \Gamma_{\delta}(\Sigma_{T^c\times T^c}). \label{AWGN:eq1*ProofOfLemmaProduct}
\end{align}
Using~\eqref{AWGN:eq1*ProofOfLemmaProduct}, Proposition~\ref{AWGN:propositionProductBound} and Proposition~\ref{AWGN:propositionInverseBound},
we obtain
\begin{align}
\left(\Sigma_{T^c\times T^c}\right)^{-1}\frac{1}{n}\sum_{k=1}^n\R{[y_{T^c,k}-G_{T^c\times \mathcal{I}}\, \mathbf{x}_k(\mathbf{w}, \mathbf{y}^{k-1})]} \in \Gamma_{\frac{\delta N^2}{\sigma_{\min}}}(I_{T^c}). \label{AWGN:eq2ProofofLemmaProduct}
\end{align}
Since
\begin{align*}
\left|\Tr\left(\left(\Sigma_{T^c\times T^c}\right)^{-1}\frac{1}{n}\sum_{k=1}^n\R{[y_{T^c,k}-G_{T^c\times \mathcal{I}}\, \mathbf{x}_k(\mathbf{w}, \mathbf{y}^{k-1})]}\right) - |T^c|\right|\le \frac{\delta N^3}{\sigma_{\min}}
\end{align*}
by~\eqref{AWGN:eq2ProofofLemmaProduct}, it follows from~\eqref{AWGN:eq1ProofofLemmaProduct} that~\eqref{AWGN:stLemmaProductProbBound} holds.

\subsection{{Derivations of \eqref{AWGN:convProofEq4*****} and \eqref{AWGN:convProofEq4****}}}\label{AWGN:appendixE}
Suppose $(\bar{\mathbf{x}}^n, \bar{\mathbf{y}}^n)$ satisfies~\eqref{AWGN:convProofEq4***}, i.e.,
\begin{align*}
K^{[\bar{\mathbf{x}}^n, \bar{\mathbf{y}}^n]} \in \mathcal{U}_{\boldsymbol{X}, \boldsymbol{Y}}^{(\delta^2, \mathbf{P}^{(\delta)})}.
\end{align*}
By the definition of~$\mathcal{U}_{\boldsymbol{X}, \boldsymbol{Y}}^{(\delta^2, \mathbf{P}^{(\delta)})}$ in~\eqref{AWGN:defSetU}, we have
 \begin{align}
 \Upsilon^{[\bar{\mathbf{x}}^n, \bar{\mathbf{y}}^n]}- \Upsilon^{[\bar{\mathbf{x}}^n,\bar{\mathbf{x}}^n]}\mathbf{G}^t&\in \Gamma_{\delta^2}(0^{N\times N}), \notag\\
   \Upsilon^{[\bar{\mathbf{y}}^n,\bar{\mathbf{x}}^n]}-\mathbf{G}\R{[\bar{\mathbf{x}}^n]}&\in \Gamma_{\delta^2}(0^{N\times N}) \label{AWGN:eq1AppendixE}
\\
\noalign{\noindent    and}
 \R{[\bar{\mathbf{y}}^n]}+ \mathbf{G} \R{[\bar{\mathbf{x}}^n]}\mathbf{G}^t - \mathbf{G} \Upsilon^{[\bar{\mathbf{x}}^n, \bar{\mathbf{y}}^n]}-  \Upsilon^{[\bar{\mathbf{y}}^n,\bar{\mathbf{x}}^n]}\mathbf{G}^t&\in \Gamma_{\delta^2}(\mathbf{\Sigma}), \notag
   \end{align}
 which implies by Proposition~\ref{AWGN:propositionProductBound} that
    \begin{align}
 \mathbf{G}\Upsilon^{[\bar{\mathbf{x}}^n, \bar{\mathbf{y}}^n]}&\in  \Gamma_{N g_{\max}{\delta^2}}\left( \mathbf{G}\R{[\bar{\mathbf{x}}^n]}\mathbf{G}^t\right),\notag\\
   \Upsilon^{[\bar{\mathbf{y}}^n,\bar{\mathbf{x}}^n]}\mathbf{G}^t&\in \Gamma_{N g_{\max}{\delta^2}}\left(\mathbf{G}\R{[\bar{\mathbf{x}}^n]}\mathbf{G}^t\right)\notag\\
\noalign{\noindent and}
 \R{[\bar{\mathbf{y}}^n]}&\in \Gamma_{(2Ng_{\max}+1){\delta^2}}\left(\mathbf{G} \R{[\bar{\mathbf{x}}^n]}\mathbf{G}^t+\mathbf{\Sigma}\right). \label{AWGN:eq2AppendixE}
   \end{align}
   Consequently, \eqref{AWGN:convProofEq4*****} and~\eqref{AWGN:convProofEq4****} follow from~\eqref{AWGN:eq1AppendixE} and~\eqref{AWGN:eq2AppendixE} respectively.

\section*{Acknowledgments}
We would like to thank the anonymous reviewer for his/her careful reading and valuable comments which have helped us greatly improve the presentation of this work.

\end{document}